# A Signal Extraction Approach for Remote Heart Rate Variability Assessment Using Proxy Measure in a Driving Simulator

Đorđe D. Nešković*[1,2], Nadica Miljković[1,3]

1. University of Belgrade - School of Electrical Engineering
2. Vinča Institute of Nuclear Sciences - National Institute of the Republic of Serbia, University of Belgrade
3. Faculty of Electrical Engineering, University of Ljubljana[1]

*Corresponding Author, e-mail: djordjeneskovic8@gmail.com

## Abstract

This study evaluates remote Photopletismography (rPPG) algorithms, Spatial Subspace Rotation (2SR), Chrominance-based method (CHROM), Plane-Orthogonal-to-Skin (POS), and Principal Component Analysis (PCA), applied to selected superpixel-based facial regions (with target counts of 10 and 20 regions) for monitoring in a driving simulator. Two novel peak enhancement approaches, based on the *Lp* norm and Fractional-Order Derivative (FOD), are introduced to enable robust Heart Rate Variability (HRV) estimation. A signal-to-noise ratio-based quality assessment of 20 s segments serves as a data cleaning mechanism to mitigate motion artifacts inherent to dynamic recording conditions. In a sample of 29 participants recorded during baseline and driving simulation conditions, Pulse Rate (PR) is calculated with clinically acceptable accuracy across configurations (validated against simultaneous Electrocardiography (ECG) recordings), achieving the lowest Mean Absolute Error (MAE) of 1.92 ± 1.72 bpm using 2SR with FOD and 20 superpixel regions. The best-case MAE reached 0.061 s for Standard Deviation of Normal-to-Normal intervals (SDNN) and 0.081 s for Root Mean Square of Successive Differences (RMSSD), with inter-beat interval detection yielding an F1 score of 0.93. Optimal parameters clustered around $p = 6–7$ for *Lp* norm and fractional orders of 1.0–1.4. All rPPG-derived parameters reproduced the statistical structure of the reference ECG across conditions and configurations. Caution is advised when using FOD due to slow changes in the rPPG waveform. Overall, 2SR is recommended for PR, while CHROM for HRV estimation, using *Lp* norm with 20 superpixels, providing clear methodological guidance for rPPG monitoring in driving simulators.

## 1. Introduction

Conventional physiological monitoring of Heart Rate Variability (HRV) parameters typically relies on contact-based sensors, including finger or ear-clip PhotoPlethysmoGraphy (PPG) [1], and wrist-worn wearable devices such as smartwatches and fitness trackers [2-3]. These approaches offer well-established accuracy and have been validated against clinical-grade reference measurements, making them the current standard for HRV assessment in a dynamic environment [2, 4-5]. However, contact-based sensors are not without limitations in the context of Driving Simulation (DS). Electrode placement and sensor attachment can interfere with the natural driving posture, while prolonged use may cause skin irritation or discomfort, potentially influencing the very physiological responses under investigation [6]. Furthermore, the setup procedure is time-consuming and requires trained personnel, limiting scalability in large-scale studies. Wrist-worn wearables, although more convenient, are susceptible to motion artifacts introduced by steering wheel interaction and are generally less accurate for HRV parameter estimation compared to chest-based Electrocardiography (ECG) recording [6]. These constraints motivate the exploration of fully contactless alternatives capable of providing comparable physiological information without physical interaction with the participant.

[1] As of January 1, 2026, Nadica Miljković is no longer affiliated with the Faculty of Electrical Engineering, University of Ljubljana. Nadica Miljković was affiliated with the University of Ljubljana at the beginning of this study.

Based on just one video, it is possible to determine a whole series of physiological parameters such as HRV, Blood Oxygen Saturation (SpO2) or Respiratory Rate (RR) simultaneously [7, 8-9]. Contactless physiological monitoring from facial video is grounded in the remote PhotoPlethysmoGraphy (rPPG), which can exploit subtle periodic variations in skin color caused by changes in blood volume associated with each cardiac cycle [10-13]. These color fluctuations, indiscernible by the naked eye, are captured by a standard RGB (Red, Green, and Blue) camera and extracted through signal processing techniques applied to selected facial regions of interest. The primary advantage of this approach lies in its fully unobtrusive nature, *i.e.*, does not include physical contact of the measurement device with the participant, and in enabling continuous monitoring over extended periods [14]. However, despite its appeal, rPPG signal extraction is inherently challenging due to the low amplitude of the physiological signal relative to numerous confounding sources, including head motion, changes in ambient illumination, specular reflections, and facial expressions, all of which introduce noise that can obscure the cardiac component [15]. To address these challenges, a range of algorithmic approaches have been proposed. Early methods relied on Independent Component Analysis (ICA) and Principal Component Analysis (PCA) to separate the rPPG signal from motion and illumination artifacts [11-12, 16]. Subsequently, model-based approaches such as Chrominance-based method (CHROM) [16-18], Plane-Orthogonal-to-Skin (POS) [16-17], and Spatial Subspace Rotation (2SR) [16, 19] were developed, exploiting assumptions about the spectral and chromatic properties of skin reflectance to improve robustness under varying conditions. Apart from the mentioned rPPG methods, techniques based on Eulerian Video Magnification (EVM) also gained considerable attention in literature [20-21]. However, EVM can be computationally intensive and equally successful results can be achieved without it [22]. Parallel efforts were focused on further improvements, such as adaptive Region of Interest (ROI) selection [23] and signal quality assessment [24] in relation to measurement conditions [25]. For ROI definition, SuperPixel (SP) segmentation has emerged as a promising method for decomposing the facial region into spatially compact sub-regions, enabling identification of the most informative areas for rPPG signal extraction [26].

In this work, we go beyond the Pulse Rate (PR) estimation within driving simulator and extend the rPPG-based analysis to a set of HRV parameters, specifically, Standard Deviation of Normal-to-Normal intervals (SDNN) [2, 3, 26-28] and Root Mean Square of Successive Differences (RMSSD) [24, 27-30], as well as the beat-to-beat alignment between individual R-peaks from the reference ECG and the corresponding peaks extracted from the rPPG signal. This multi-level evaluation, encompassing PR estimation, rPPG-based HRV parameter accuracy, and individual peak correspondence, provides a more complete picture of the physiological information that can be reliably retrieved from a facial video. To this end, we utilize a unique dataset comprising 29 participants and a wide range of PR, reflecting the individual physiological diversity encountered in real-world driving. In addition to four established rPPG algorithms, CHROM [16-18], POS [16-17], 2SR [16, 19], and PCA-based approach [11-12, 16], we introduce two peak enhancement techniques, namely the *Lp* norm [31] and the Grünwald–Letnikov Fractional-Order Derivative (GL FOD) [32-33], designed to enhance the peaks originating from periodic color changes induced by pulsatile blood volume variations in the cutaneous capillaries of the facial skin. In addition to comprehensively evaluating the accuracy of PR, SDNN, and RMSSD across all algorithm configurations and recording conditions, we employ signal-to-noise ratio (SNR) estimation to automatically select high-quality 20-second segments with minimal motion artifacts. The key objective of the study is to confirm whether the observed changes in autonomic activity are detectable through rPPG-based estimation. All extracted parameters are evaluated against a simultaneously recorded reference ECG signal, which serves as the ground truth.

## 1.1. Research Questions

Altogether, this research goes beyond simple PR estimation, by introducing Inter-Beat-Interval (IBI), SDNN, and RMSSD calculation, which are sensitive markers of parasympathetic nervous system activity, enabling a broader assessment of autonomic nervous system activity from facial video [27]. The goal is not only to determine whether camera-based HRV monitoring can serve as a viable alternative to contact-based physiological sensing in realistic driving research settings, but also to provide practical guidance for researchers considering the adoption of rPPG-based physiological monitoring in driving simulators — specifically, which combination of rPPG method, peak enhancement strategy, and spatial configuration yields the most reliable estimates of PR and HRV parameters under dynamic and static conditions. Specific research questions are:

1. Which ROI determined by SP contributes most effectively to estimation of rPPG-based HRV parameters in a simulator environment? For this purpose, we use Mean Absolute Error (MAE), between HR from reference ECG and estimated PR from the video, as well as estimated SNR.
2. In the context of a driving simulator environment, which rPPG approach—PCA, CHROM, POS, or 2SR—provides the most precise extraction of rPPG-based HRV parameters (PR, SDNN, and RMSSD) when compared to the reference ECG? Moreover, how accurately can rPPG-based HRV parameters resemble reference ECG measurements during Baseline (BSL) and DS conditions?
3. How does peak enhancement achieved through *Lp* norm–based processing and GL FOD affect the accuracy of all extracted rPPG-derived HRV parameters?
4. How reliably can peaks derived from rPPG signals be identified in relation to the reference R-peaks from ECG recordings during BSL and DS conditions?

# 2. Method

Complete analysis is employed using the Matlab version 2023a (The MathWorks, Natick, USA). The proposed processing pipeline is illustrated in Fig. 1. The study is conducted in a motion-based DS at the Faculty of Electrical Engineering, University of Ljubljana, featuring authentic vehicle controls (*e.g.*, seat, steering wheel, and pedals) and with a physical dashboard. Visuals are displayed on three curved HD screens (1920×1080 pixels) providing a 145° field of view. The scenario, simulating highway driving, is developed in SCANeR Studio [34]. The protocol adhered to the Ethical guidelines of the University of Ljubljana (approval No. #049-2025) and to the Declaration of Helsinki. All participants signed written Informed Consents.

## 2.1. Recording Procedure

Participants completed a ~30-minute session comprising three phases: a 15-minute seated BSL period, a brief familiarization with the simulator controls, and a ~12-minute DS phase consisting of highway driving under normal conditions followed by a sudden dense fog episode designed to elicit physiological arousal. Facial video was recorded continuously throughout all phases using an ELP HD camera (1920×1080 px, 30 fps), synchronized with a reference ECG signal acquired at 1000 Hz via the PLUX device (PLUX Wireless Biosignals S.A., Lisbon, Portugal). Recording sessions were conducted under controlled illumination conditions (743.08 ± 10.31 lx). A total of 29 participants (17 males, 12 females; age 25.86 ± 5.75 years, range 21–42) completed the protocol. Full details of the recording procedure are provided in the Supplementary Materials [35].

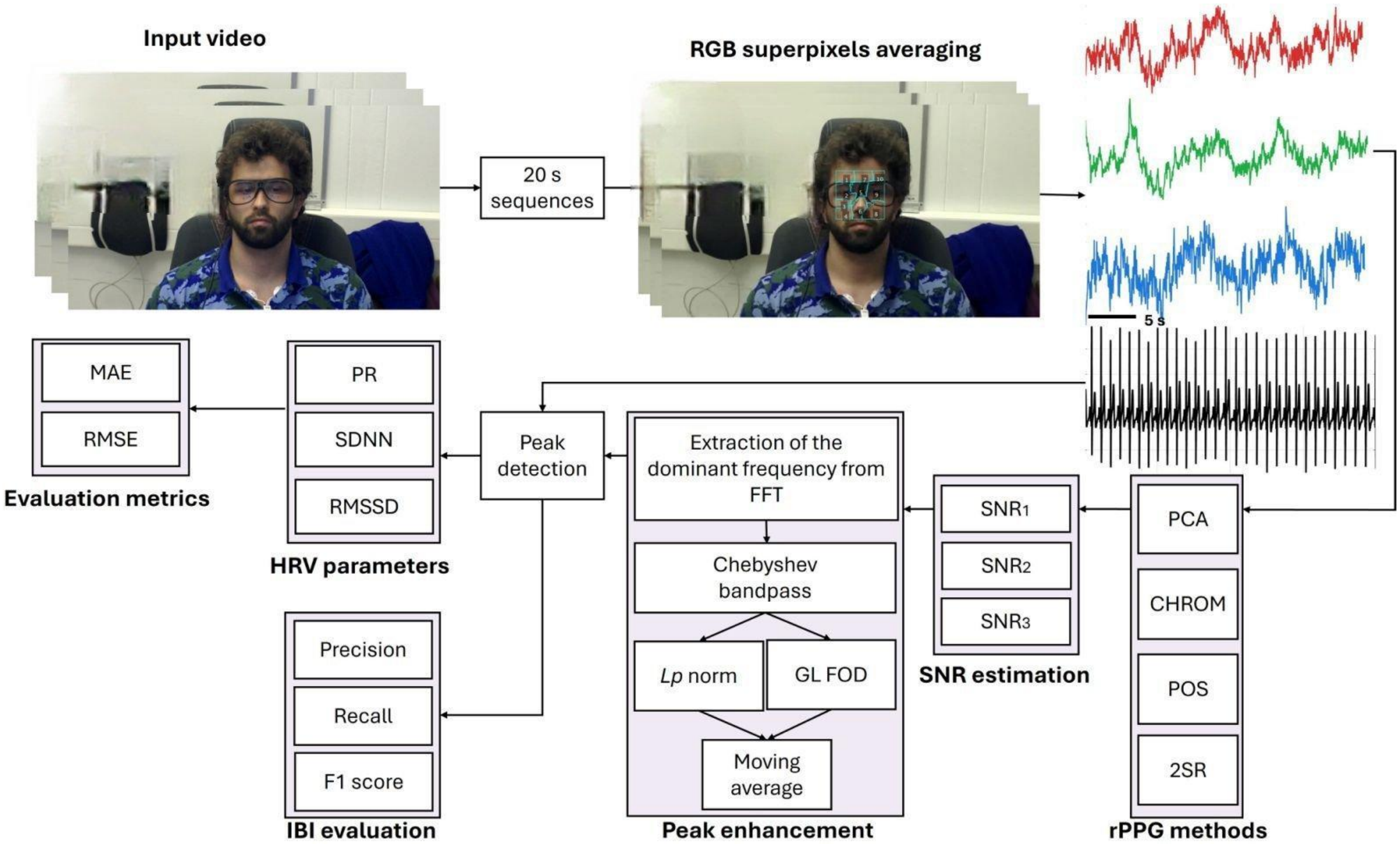

Figure 1: Block diagram of proposed method for rPPG performance. Abbreviations are: Red Green Blue (RGB), Pulse Rate (PR), Heart Rate Variability (HRV), Standard Deviation of Normal-to-Normal intervals (SDNN), Root Mean Square of Successive Differences (RMSSD), Mean Absolute Error (MAE), Root Mean Square Error (RMSE), Signal-to-Noise Ratio (SNR), Principal Component Analysis (PCA), Chrominance-based method (CHROM), Plane-Orthogonal-to-Skin (POS), Spatial Subspace Rotation (2SR), Grünwald–Letnikov Fractional Order Derivative (GL FOD), Fast Fourier Transform (FFT), and Inter-Beat-Interval (IBI).

## 2.2. Superpixel-based Selection of Regions of Interest

For each 20 s segment, the ROI is defined through a hierarchical localization pipeline consisting of coarse head detection, refinement to the facial skin region, and subsequent decomposition into SP-based sub-regions. Head detection is performed using a Viola-Jones cascade detector applied to the first frame, with false positive bounding boxes eliminated via a minimum size threshold of 150 pixels, empirically determined based on the recording resolution and participant framing. The detected bounding box is geometrically refined through symmetric horizontal insets and asymmetric vertical trimming to exclude the hairline, ears, chin, and neck, centering the ROI on the mid-face area of high skin perfusion. The refined region is further decomposed into SP-based sub-regions using the simple linear iterative clustering algorithm, with target counts of 10 and 20 SP (previously explored in pilot tests) selected as a compromise between computational complexity and spatial representativeness. To maintain spatial correspondence across frames in the presence of head movement, an optical flow-based Lucas-Kanade tracker warps the SP polygon boundaries to align with the facial geometry in each subsequent frame. Full details of the localization and tracking procedure are provided in the Supplementary Materials [35].

In each segment, facial skin regions are determined via SP-based segmentation. For each SP - defined region and for each frame two types of signals are extracted per region. First, mean RGB values are computed within each polygon, yielding three-dimensional color time series per SP (used for CHROM [16-18], POS [16-17], and PCA (where the first Principal Component (PC1) is selected) [11-12, 16] rPPG methods). Second, the spectral subspace decomposition required for the 2SR method [16, 19] is derived from the covariance structure of pixel colors within each region, producing a unitary basis matrix and a corresponding set of eigenvalues at each frame. These quantities together constitute the spatiotemporal data matrix used for subsequent rPPG estimation. Fig. 2 presents the rPPG signals obtained using four different methods: 2SR, CHROM, POS, and PCA. For visualization purposes, only zero-phase filtering applied in both forward and reverse directions is performed using the fourth-order Chebyshev Type II bandpass filter with a frequency range of 0.7 Hz – 3 Hz.

## 2.3. Selection of Optimal SuperPixels and Useful Segments

The recorded facial videos obtained during DS or BSL conditions are partitioned into consecutive 20-second long segments at the outset without overlapping [26]. Variable window lengths are not examined in detail in this study, instead we follow prior recommendations, and pilot tests, to employ 20 s segments [23, 26] as an initial step of the processing pipeline (Fig. 1). Certain segments are excluded from further analysis due to failed initial face detection, caused by head rotations or hand-to-face occlusions at the beginning of the 20 s long segment. Such occurrences are inherent to the dynamic nature of a DS environment. Segmentation permits independent extraction and analysis of rPPG signals from shorter time windows, following prevalent protocols in HRV assessment via remote PPG [23, 26]. We compare SP-based selection of regions to evaluate which regions are the most suitable for pulse extraction, and the execution time for each case is measured.

A single SP region is selected as optimal using a two-stage criterion combining estimated SNR and accuracy of PR detection (*i.e.,* MAE). The SNR is calculated following the application of the specific rPPG extraction method (CHROM, PCA, POS, or 2SR), while the MAE is determined based on processed signals through the peak enhancement process, as illustrated in Fig. 1. To estimate SNR (in three different ways), we compute the ratio of the power of the useful signal component to the noise component, following the recommendation in [36], where a similar two-stage approach by MAE and SNR application is used for the proper region selection for rPPG calculation. We adopt a two-stage criterion to ensure both accurate PR detection and minimal noise influence—selecting the region with the lowest MAE while also favouring the least noise contamination (*i.e.,* the highest SNR).

First, all candidate regions, each individually, are evaluated using the MAE between the rPPG-derived PR and the reference ECG-based Heart Rate (HR) for BSL and DS separately. Relatively high SNR alone may not guarantee accurate pulse estimation in rPPG signals, since periodic components unrelated to cardiac activity (*e.g.*, head or body motion during the DS) can produce dominant peaks in the Power Spectrum Density (PSD) within the typical rPPG frequency band (0.7 Hz – 3 Hz), resulting in artificially high SNR values despite not corresponding to the true HR. A similar limitation of SNR-based evaluation has been reported by Benezeth *et al.* [37], where the Authors emphasized that SNR does not account for the temporal morphology of the rPPG signal, which is an important indicator of signal quality. Therefore, we use SNR as a secondary criterion. Firstly, the four regions exhibiting the lowest MAE values (primary criterion) are retained as physiologically plausible candidates, where MAE is computed in the time domain as the mean absolute difference between the PR estimated from the peaks detected in the extracted rPPG signal and the reference HR derived from the ECG signal, both evaluated over each 20 s segment independently. This initial selection step aims to exclude regions that produce physiologically incorrect periodic components. Then, among these candidate regions, the final selection is performed based on the highest SNR averaged across all 20-seconds long sequences for BSL and DS regimes. The secondary SNR-based criterion is applied after MAE-based filtering to favor regions providing the most accurate and noise-robust pulsatile signal while maintaining agreement with the reference measurements.

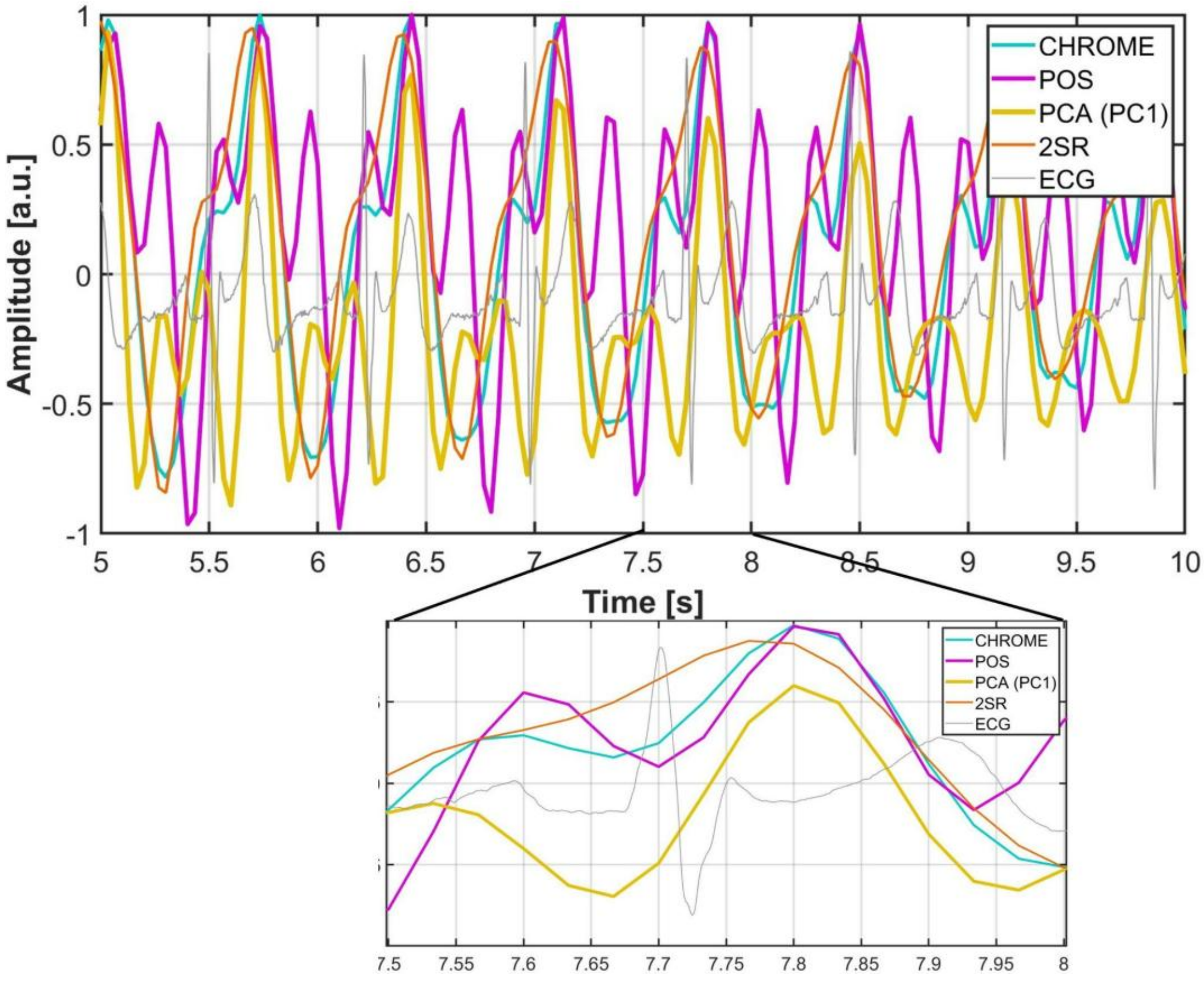

Figure 2: Display of rPPG signal extraction after applying the CHROM, PCA (the first Principal Component, PC1), POS, and 2SR methods before peak enhancement.

When recorded with a camera, PPG signal is commonly inverted in comparison to measurement obtained from contact PPG sensor [38-39]. Therefore, a pilot analysis using simulated sinusoidal RGB components at 60 bpm with mutual phase shifts [38] of ±5° is performed and it revealed that CHROM and POS produce signals with phase shifts of approximately −90° relative to the G channel, meaning that the peak is located with shift about −90°. Since such delays vary due to the changes in PR and in the presence of noise, we decide not to invert CHROM and POS. PCA closely tracks the G channel (−5° phase shift), meaning it is inverted in comparison to PPG. It should be noted that for PR, SDNN, and RMSSD calculation, phase delay does not introduce changes as it depends on the peak-to-peak distance. However, IBI detection depends on the phase delay, and we present results for IBI detection for both inverted and noninverted PCA in the Supplementary Materials [35] (Tables A6-A7). Phase delay of the 2SR method does not exist and its final output is not directly expressible through raw RGB channels. Full details and signal visualizations are provided in the Supplementary Materials (Fig. A1).

## 2.4. Signal-to-Noise-Ratio Metrics

The SNR of the extracted rPPG signal is estimated via power spectral analysis, with the PSD computed as the squared magnitude of the discrete FFT of each 20 s segment. Three distinct SNR metrics are defined, each serving a specific purpose in the pipeline. The first metric (SNR narrow filt. ECG) is computed as the ratio of PSD within a narrow band of ±0.1 Hz centered at the ECG-derived dominant HR frequency to the remaining out-of-band power, expressed in decibels (dB), following the approach of [40]. This metric is used for selecting the facial SP region with the highest signal quality.

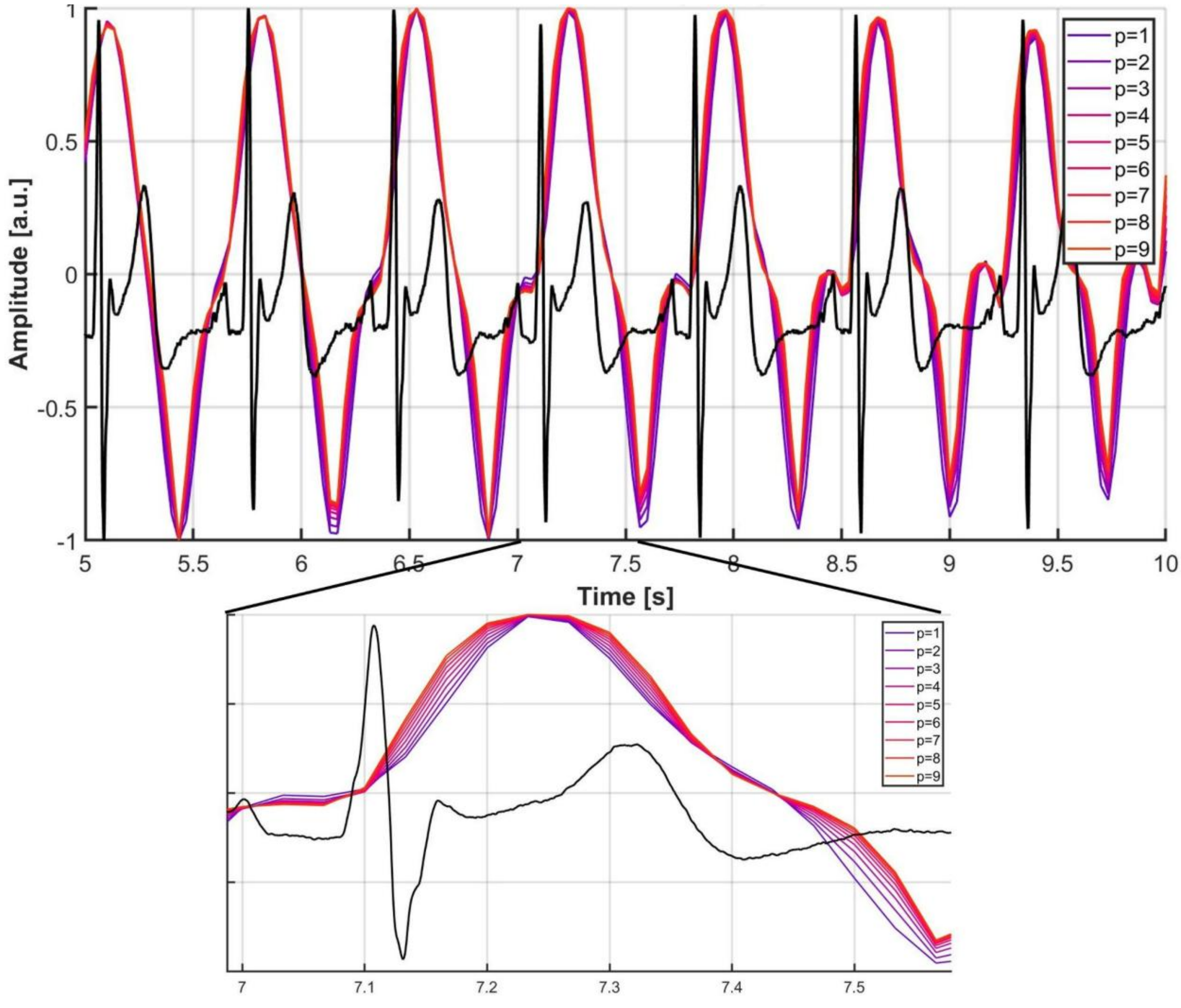

Figure 3: The impact of different $p$ values on the application of the $Lp$ norm on rPPG signal with reference ECG presented in graphs. For $p = 1$, the $Lp$ norm actually does not change the extracted rPPG signal.

The second metric (SNR 0.7–3 Hz) integrates PSD over the full physiological HR band of 0.7 Hz – 3 Hz, with a ±0.1 Hz notch at 2 Hz excluded from both signal and noise calculations to suppress a recurring non-physiological spectral artifact observed in several recordings [37, 41]. This metric serves as a segment rejection criterion, with segments falling below −17 dB discarded from further analysis [41]. Finally, the third metric (SNR narrow filt. rPPG) is computed identically to SNR narrow filt. ECG, but with the center frequency determined from the dominant spectral peak of the rPPG signal itself rather than the reference ECG. This self-referenced metric enables comparison between ECG- and rPPG-derived dominant frequencies, and is applicable in practical deployments where a simultaneous ECG reference is unavailable.

## 2.5. Peak Enhancement in rPPG

The extracted rPPG signal features undergo additional processing steps aimed at enhancing peak prominence. The proposed processing is inspired by the Pan-Tompkins algorithm for QRS detection in ECG signals [42]. To identify a narrow passband containing pulsatile information, the pulse frequency is determined as the frequency of the dominant spectral peak within the physiological HR range of 0.7 Hz – 3 Hz in the PSD of the rPPG signal (marked as the "Extraction of the dominant frequency from FFT" block in Fig. 1). Before identifying the dominant frequency, a third-order Butterworth notch filter centered at 2 Hz with a ±0.1 Hz bandwidth is applied to suppress the recurring spectral artifact. The filter is implemented using zero-phase filtering in both forward and reverse directions to avoid phase distortion. All subsequent processing steps are performed independently for each non-overlapping 20 s signal segment.

Based on the detected pulse frequency, a fourth-order Chebyshev type II bandpass filter is applied using an adaptive bandwidth determined from HRV observed in the reference ECG signal during BSL and DS conditions. Specifically, the passband width around the fundamental frequency is defined as ±3SD, where SD represents the Standard Deviation of the ECG-derived pulse from all 20 s sequences. In addition to the fundamental component, the filter passband is extended to include the first and the second harmonics [19], using bandwidths of ±4SD and ±5SD, respectively, to account for increased spectral dispersion at higher harmonics. Passband widths of ±3SD, ±4SD, and ±5SD are selected through pilot iterative testing. Zero-phase filtering is employed in both directions to ensure that the signal phase remains unaltered. The Chebyshev type II filter is chosen for its monotonic passband response and steeper roll-off, which allows more effective suppression of out-of-band noise while preserving physiologically meaningful pulsatile components. While the Butterworth filter is commonly used in biomedical signal analysis [22], Chebyshev type II has been shown to be particularly effective for denoising rPPG signals [43].

Subsequently, to examine the most appropriate method for peak enhancement, we apply either the Grünwald–Letnikov Fractional Order Derivative (GL FOD) [32-33] as presented in Equations (1) and (2) with fractional order in range from 1 to 3 (step is 0.1) or an *Lp*-norm–based peak enhancement computed using a built-in Matlab function *norm* applied over sliding windows (width is four samples, which is approximately 130 ms, with a step of one sample corresponding to 32.5 ms) of the signal feature [31]. The *Lp* norm is calculated locally within each window for $p$ values ranging from 1 to 9 by using Equation (3) to evaluate the trade-off between noise suppression at lower $p$ values and the amplification of dominant rPPG features at higher $p$ values, which is critical in motion-intensive driving scenarios (Fig. 3). In this paper, the *Lp*-norm operation, formulated as a forward sliding-window estimator, introduces a deterministic group delay equal to half of the window length. For the selected window size of four, the delay amounts to two samples (≈67 ms at 30 fps). As this delay is constant and independent of signal content, temporal compensation is applied prior to peak detection to ensure accurate alignment with detected R peaks from the reference ECG signal. Implemented approaches for peak enhancement (*Lp* norm and GL FOD) are applied independently (Figs. 3 and 4 show the effect of both enhancement methods) and subsequently compared for their efficacy in peak enhancement (Fig. 1).

$$D^{\alpha} f(t) = \frac{1}{h^{\alpha}} \sum_{k=0}^{\frac{t-a}{h}} (-1)^k \binom{\alpha}{k} f(t-kh) \tag{1}$$

$$\binom{\alpha}{k} = \frac{\Gamma(k+1)}{\Gamma(\alpha-k+1)\Gamma(\alpha+1)} \tag{2}$$

$$L_p^{norm}(n) = (\frac{1}{Nw}) \sum_{i=n}^{n+Nw-1} |x^p|)^{\frac{1}{p}} \tag{3}$$

Before threshold application for rPPG peak detection, the signal smoothing is applied using a weighted Moving Average (MA) filter based on a normalized Hann's window [18] of width 12 samples (400 ms) as proposed in [22]. The filter is applied using zero-phase filtering in both directions, ensuring amplitude preservation and preventing phase distortion that could affect peak localization. Finally, the *findpeaks* function is employed for the exact peak localization with the minimum peak distance set to 0.33 s, reflecting the assumption that, under DS conditions in a seated position, healthy subjects are unlikely to exhibit pulse rates exceeding 180 bpm. To improve robustness, application of a threshold-based procedure relied on the prominence criterion rather than on a fixed amplitude threshold, allowing reliable identification of both dominant physiological peaks and smaller peaks potentially associated with noise. Consequently, a fixed height-based threshold would be unsuitable, as it could either suppress relevant peaks or admit excessive noise.

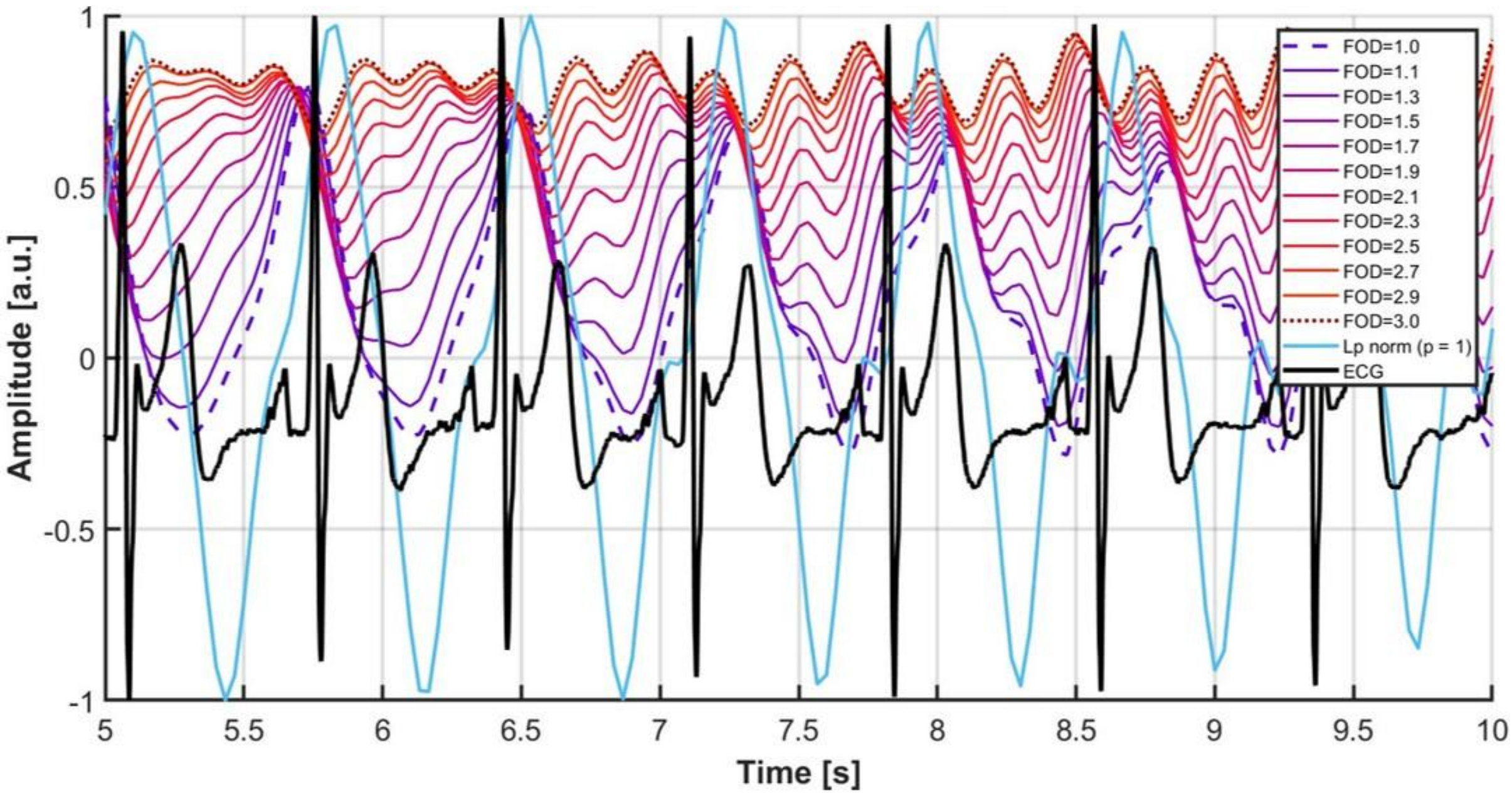

Figure 4: The impact of different values of the Fractional Order Derivative (FOD) for the application of GL FOD for peak enhancement in rPPG signal with presented reference ECG. The value of FOD of 1 is equal to the application of standardized Pan-Tompkins algorithm. Also, the extracted rPPG without peak enhancement is used. A notable change in peak location is noticed due to the slow changes of rPPG signal, making FOD a questionable approach for inter-beat interval detection.

A minimum prominence threshold of 0.05 is applied to filter out residual noise-induced fluctuations. This threshold defines the minimum required amplitude difference between a detected peak and its local baseline, as implemented via the *MinPeakProminence* parameter of the Matlab *findpeaks* function. The value of 0.05 is determined empirically as a compromise between suppressing spurious detections and preserving valid cardiac peaks.

## 2.6. Evaluation of Heart Rate Variability Parameters

Based on detected peaks in the time domain [44], the success of physiological parameter extraction is evaluated and compared against the reference ECG signal. The accuracy of extraction of the relevant rPPG-based HRV parameters (PR, SDNN, and RMSSD [27, 30]) is evaluated using the MAE and the Root Mean Square Error (RMSE). Following the rPPG peak detection process, the PR for each 20-second segment is derived by calculating the median of the inter-peak intervals. The final PR estimate is then computed as the inverse of this median value, scaled to bpm. In line with the findings in [27, 29], which suggest that at least 20 s of data are required to obtain meaningful HRV indices, both rPPG-based and reference ECG parameters are extracted from 20-seconds long segments to ensure statistical validity.

## 2.7. Inter-Beat-Intervals Evaluations

rPPG pulse peak follows an ECG R-peak due to the physiological delay required for the pulse wave to propagate from the heart to the periphery [45-46]. According to empirical data from [47-48], Pulse Transit Time (PTT) for peripheral sites typically falls within the range of approximately 0.1 s to 0.6 s. Specifically, based on the measurement site, PTT is expected to be approximately 150 ms when measured on the ears, 250 ms on the fingers, and 350 ms on the toes [47-48]. Due to the temporal and physiological consistency, detected rPPG peaks could be classified as True Positives (TP), True

Negatives (TN), False Positives (FP), or False Negatives (FN), allowing performance evaluation using classification metrics, Precision, Recall, and F1 score, with 150 ms tolerance in relation to the identified R-peak in ECG signal [49-50]. Defining a strict PTT limit for the facial rPPG is not straightforward, as PTT is known to vary across individuals due to their height, blood pressure, or arousal level [47-48], but also the blood does not pulse synchronously on the face regions [51].

Therefore, three evaluation criteria are considered to assess rPPG peak detection performance. For the first evaluation criteria (Strict), a more stringent case, an rPPG peak is classified as a TP if it occurs after the corresponding ECG R-peak within a temporal window of 250 ms. Although physiologically expected PTT values are closer to 150 ms for proximal measurement sites, temporal uncertainty arises from differences in signal resolution, as ECG signals are sampled at 1000 Hz, whereas video recordings are acquired at 30 fps, limiting temporal precision in rPPG peak localization. Additionally, the temporal window is extended to 350 ms (Medium) to account for potential increases in PTT and further decrease in IBI metrics caused by lower frame rates in comparison to the sampling frequency of reference ECG. An rPPG peak detected between two consecutive R-peaks, but outside the 250 ms interval following the first R-peak is classified as an FP. An FN is assigned when no rPPG peak is detected between two consecutive ECG R-peaks. For the third evaluation criteria (Relaxed), we use a more relaxed threshold, where temporal alignment constraints are reduced. A TP is defined when exactly one rPPG peak is detected between two consecutive ECG R-peaks. An FP is assigned when two or more rPPG peaks are detected within the same R-R interval, while an FN corresponds to the absence of any rPPG peak between consecutive R-peaks. A similar dual-criterion approach based on peak-to-peak comparison using narrower and wider temporal tolerance windows has previously been applied demonstrating the usefulness of multi-level evaluation criteria when temporal alignment uncertainty is present [32].

The TP, FP, and FN counts are first determined for each individual 20-second segment across both evaluation criteria. These counts are then accumulated across all segments within a recording, and the overall Precision, Recall, and F1 score are computed from the cumulative totals, yielding a single set of performance metrics per recording that reflects the global peak detection accuracy over the entire session. To evaluate performance, IBI values are extracted using all implemented rPPG methods (CHROM, PCA, POS, and 2SR) across two distinct spatial configurations, initially employing 10 and 20 SP regions. This approach allows us to evaluate whether increasing the number of SP regions from 10 to 20 enhances rPPG signal extraction quality, while simultaneously quantifying the associated computational trade-offs in processing time. Furthermore, computational efficiency is quantified by measuring the elapsed time for code execution of each configuration. Elapsed time is then averaged across 20 s long sequences for both BSL and DS.

## 2.8. Peak Detection in Reference ECG

The reference ECG signal is segmented into non-overlapping 20 s sequences to match the rPPG segmentation. R-peak detection is performed using a modified Pan–Tompkins algorithm [52], incorporating bandpass filtering (0.5 Hz – 20 Hz, fourth-order Butterworth), differentiation, squaring, and a 250 ms Hann-windowed moving average envelope. Rather than applying a fixed global threshold, an adaptive threshold set to half of the third quantile of candidate peak amplitudes is employed, improving robustness in noisy recordings and preventing pathological beats with exceptionally high amplitudes from suppressing the detection of subsequent normal beats [53]. Full methodological details are provided in the Supplementary Materials [35].

## 2.9. Statistical Analysis

Statistical analysis is performed to evaluate whether significant differences exist between BSL and DS periods for both ECG-derived reference parameters and rPPG-extracted parameters, across all evaluated configurations. Normality is assessed using the Shapiro-Wilk test. Normally distributed samples are compared using a two-sample Student's t-test with Cohen's d as the effect size measure, while non-normally distributed samples are compared using the Mann-Whitney U test with Cliff's $\delta$. A significance threshold of $\alpha = 0.05$ is adopted throughout. Effect size is additionally computed as a complementary measure of practical relevance, given that statistical significance alone does not preclude the possibility of trivially small differences being declared significant when sample sizes are sufficiently large [54]. The analysis is first applied to the ECG-derived PR, SDNN, and RMSSD as a reference benchmark, and subsequently repeated for all rPPG configurations. This systematic evaluation allows for a direct assessment of whether a given rPPG processing configuration is capable of capturing the physiologically meaningful cardiovascular changes induced by the DS task, in a manner consistent with the reference ECG-based findings.

# 3. Results

The average execution time per 20 s signal segment is 178±35 s when using the 20 SP configuration, whereas the 10 SP configuration reduces this time to 96± 17 s. Considering that each recording consists of approximately 70 segments (each 20 s in duration), the total average processing time per participant amounted to approximately 3 hours and 30 minutes for the 20 SP approach, and approximately 1 hour and 50 minutes for the 10 SP approach.

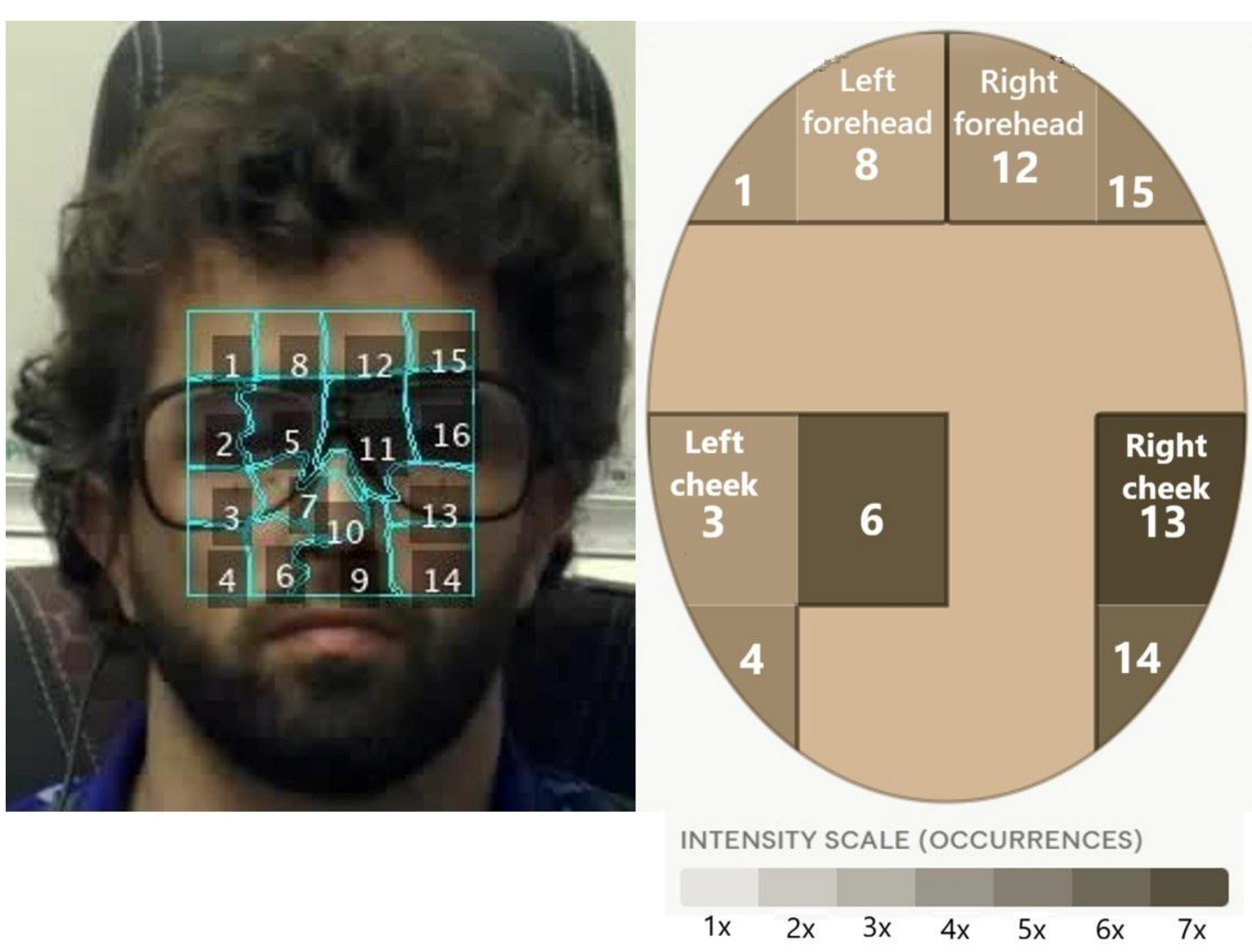

Figure 5: Example of regions that proved to be the most useful for extracting the rPPG signal in one subject when 20 SP are initially determined. A concrete example refers to the *Lp* norm and the 2SR method. The participant whose facie is shown provided informed consent for the use of their image in this publication.

The data cleaning process revealed that the primary reason for segment exclusion is a failure in face detection, with most recordings losing between 6 and 10 sequences (out of approximately 70 per participant). The incidence of detection failure ranged from zero in the most stable recordings to a maximum of 68 out of 74 segments in a single case where the participant maintained a persistently lowered head position. Additionally, the rejection of segments based on the SNR threshold varied significantly across the applied algorithms. The PCA method proved the least robust in this regard, with approximately 20 segments discarded per recording, whereas the other methods demonstrated higher signal stability with only zero to five rejections.

An illustrative example of the regions that contribute most to rPPG signal extraction for a single subject is provided in Fig. 5, considering the case of 20 initially defined SP regions using the *Lp* norm approach in combination with the 2SR method. A broader overview of the most informative regions across both configurations of initially defined SP regions, including both peak enhancement approaches and all four rPPG methods, is given in Table A1 (in Supplementary Materials [35]). For the initially defined 20 regions, when applying the GL FOD approach, the most frequently selected fractional orders that proved to be optimal are 1 for the 2SR method, 1.4 for CHROM, 1.1 for PCA, and 1 for POS. For the case of 10 initially defined SP regions, the corresponding optimal values are 1.2, 1.4, 1.1, and 1.1 for 2SR, CHROM, PCA, and POS, respectively. When applying the *Lp* norm, the most frequently observed optimal values of the parameter $p$ for the 20 initial regions are 7, 6, 6, and 5 for 2SR, CHROM, PCA, and POS, respectively. For the 10 initial SP regions, these values are 7, 9, 6, and 6, respectively.

Figs. 6 with Figs. A9, and A10 (in the Supplementary Materials [35]) present the results obtained for three different SNRs. The results are shown for both peak enhancement approaches (*Lp* norm and GL FOD) in combination with the four rPPG methods (2SR, CHROM, POS, and PCA) and correspond to the initially defined 20 SP regions. Table A5 summarises all of three different SNR values for initially defined 10 SP regions [35].

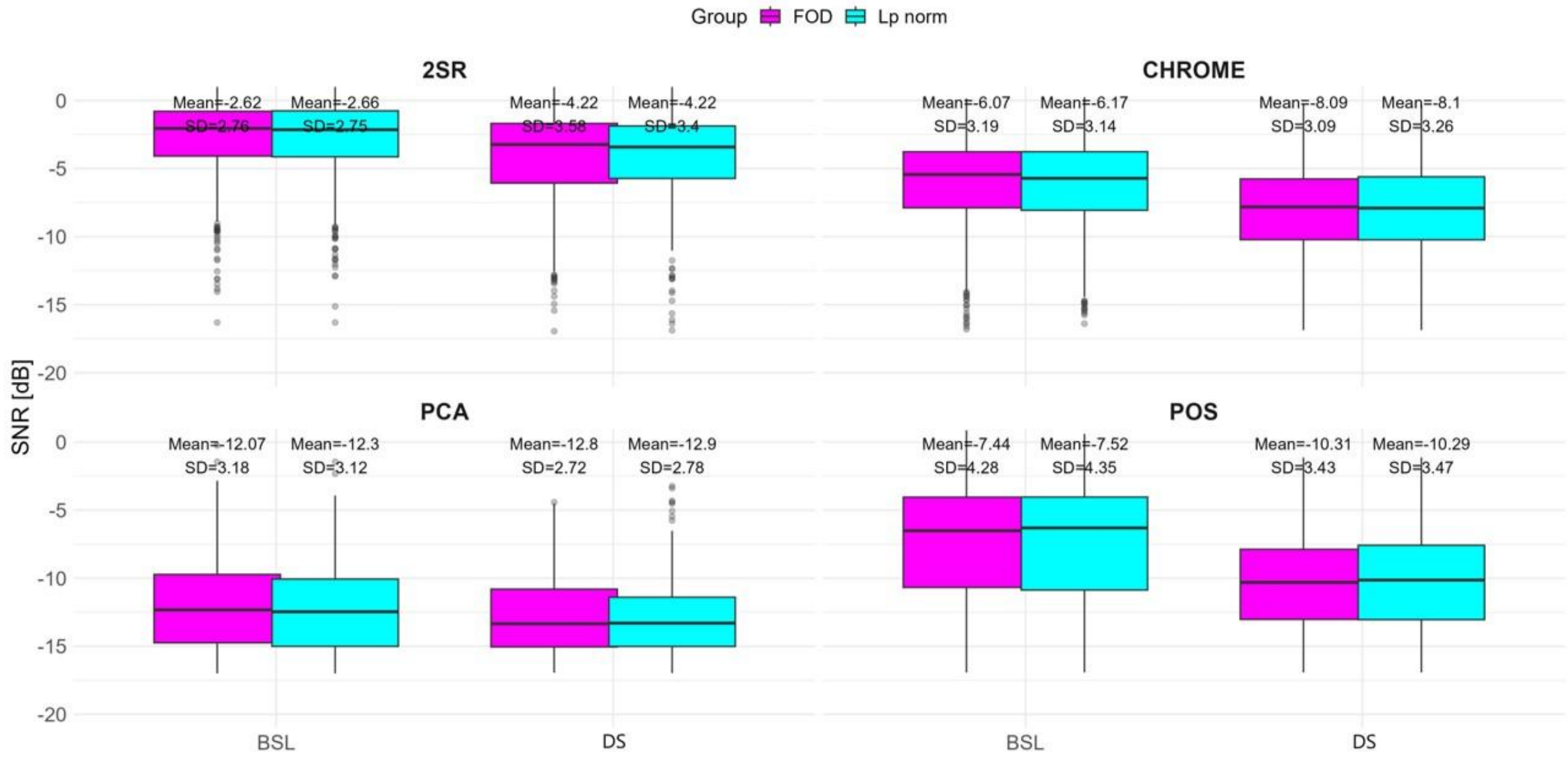

Figure 6: Comparison of SNR values (useful part of PSD defined within a band range from 0.7 Hz to 3 Hz) across two peak enhancement approaches and four rPPG methods. Shown graphic corresponds to initially determined 20 SP.

Fig. 7 shows the Bland–Altman plots comparing the reference HR obtained from ECG with the extracted PR values using the two proposed peak enhancement approaches (*Lp* norm and GL FOD) in combination with four rPPG methods (2SR, CHROM, POS, and PCA) for the initially defined 20 SP regions. The mean value of the reference HR is 71.06±3.19 bpm during BSL and 81.52±4.90 bpm during DS. The quantitative evaluation is further presented in Table A2 (Supplementary Materials [35]) for both 10 and 20 initially defined SP regions. Specifically, Fig. 8 depicts the MAE between the reference HR and the extracted PR values,

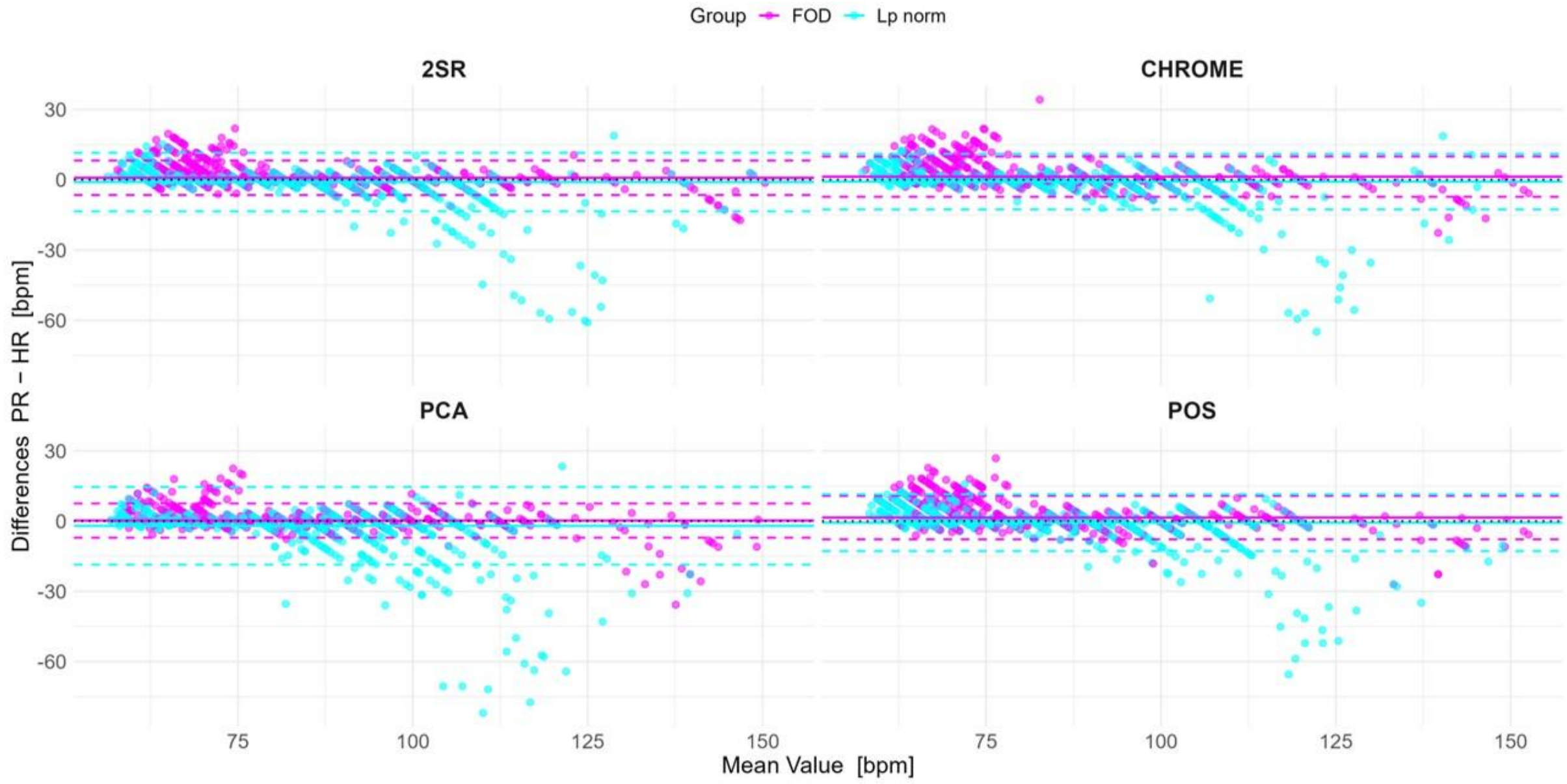

Figure 7: Bland-Altman plots for two peak enhancement approaches per each rPPG method. Shown graphic corresponds to the initially determined 20 SP.

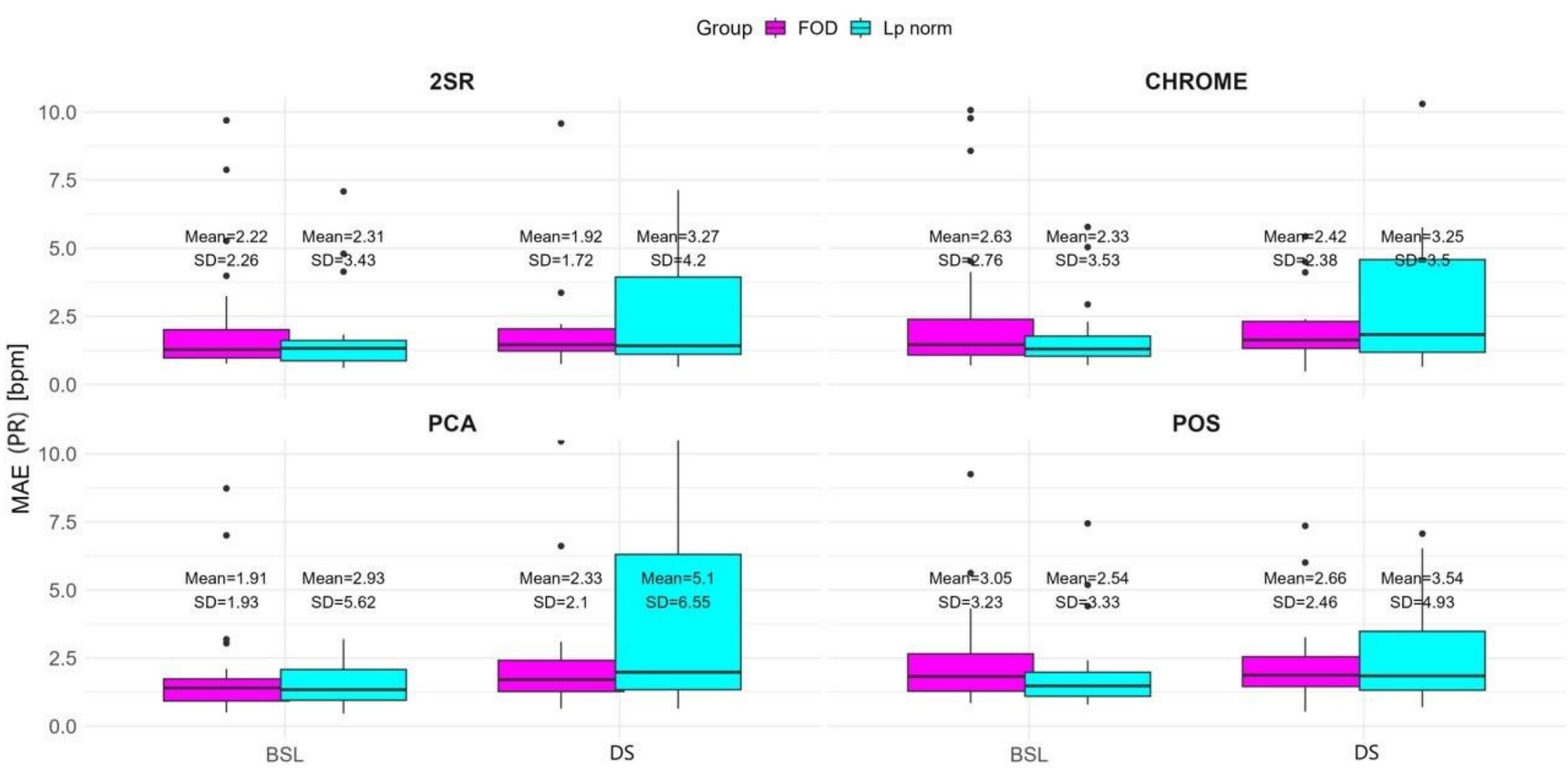

Figure 8: MAE between HR from the reference ECG and extracted PR for two peak enhancement approaches per each rPPG method. Shown graphic corresponds to initially determined 20 SP.

Reference ECG data imply that mean SDNN is 0.044±0.036 s during BSL and 0.43±0.035 s during DS. Fig. 9 and Table A3 (in Supplementary Materials [35]) present the MAE and RMSE, respectively, between the reference SDNN values obtained from ECG and the SDNN values extracted using the proposed methods. The results shown in Fig. 9 correspond to the initially defined 20 SP regions, while results of all four rPPG methods, both SP configuration during BSL and DS are summarized in Supplementary Materials in Table A3 [35]. According to the reference ECG data, mean RMSSD during BSL is 0.044±0.040 s, while mean RMSSD during DS is 0.045±0.045 s. Fig. 10 and Table A4 in Supplementary Materials [35] illustrate the MAE and RMSE, respectively, calculated between the reference RMSSD values derived from ECG and the RMSSD values obtained using the two peak enhancement approaches and four rPPG methods. The results presented in these figures refer to the initially defined 20 SP regions, while a complete overview of results for each rPPG method, both SP configuration during BSL and DS are provided in Supplementary Materials in Table A4 [35].

Tables 1, A6, and A7 from Supplementary Materials [35] report the IBI-based evaluation metrics, including Precision, Recall, and F1 score. These results are presented for three different evaluation criteria, namely the Strict, Medium, and Relaxed approaches used for metric calculation.

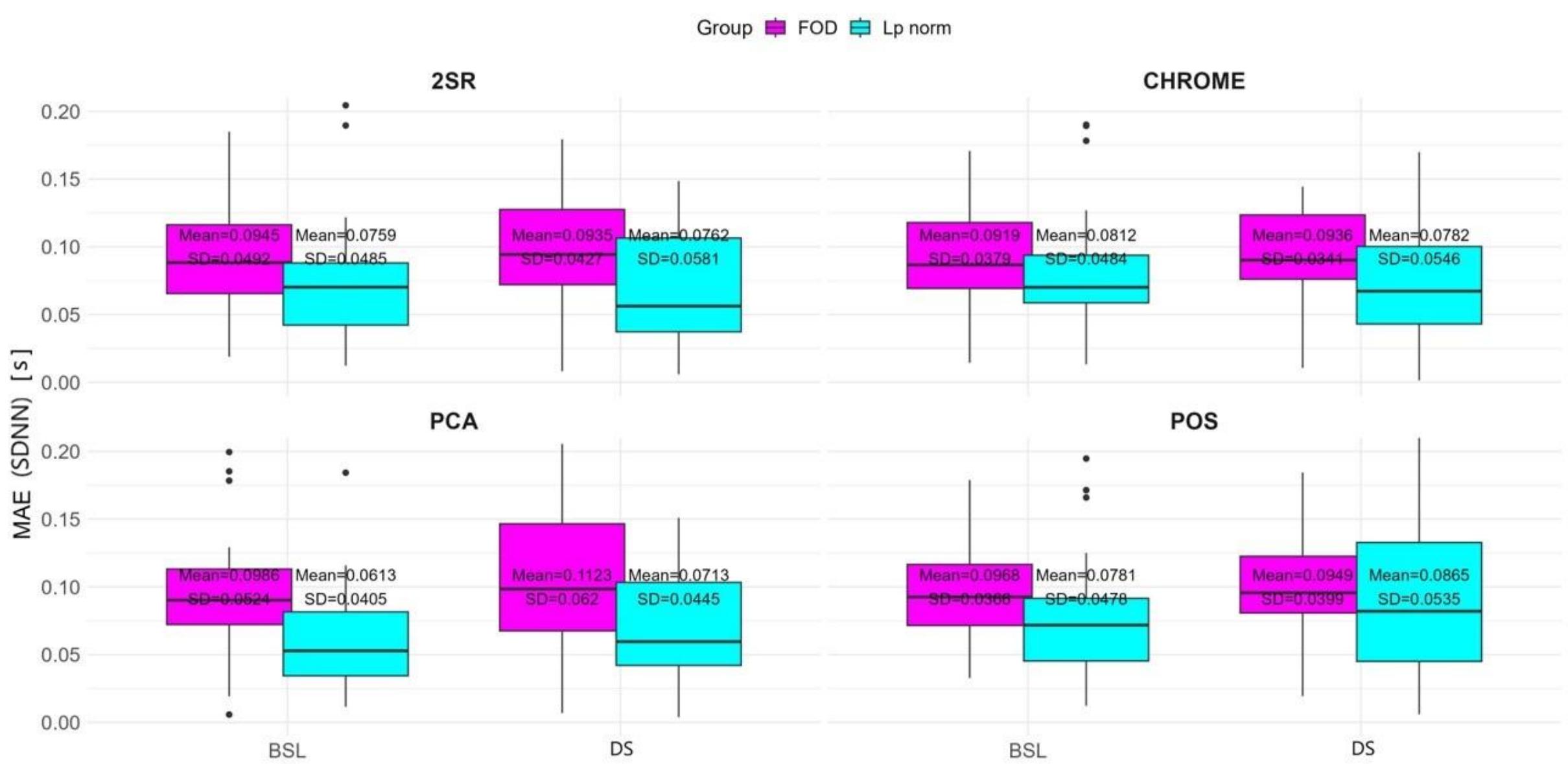

Figure 9: MAE between SDNN from the reference ECG and extracted SDNN for two peak enhancement approaches per each rPPG method. Shown graphic corresponds to initially determined 20 SP.

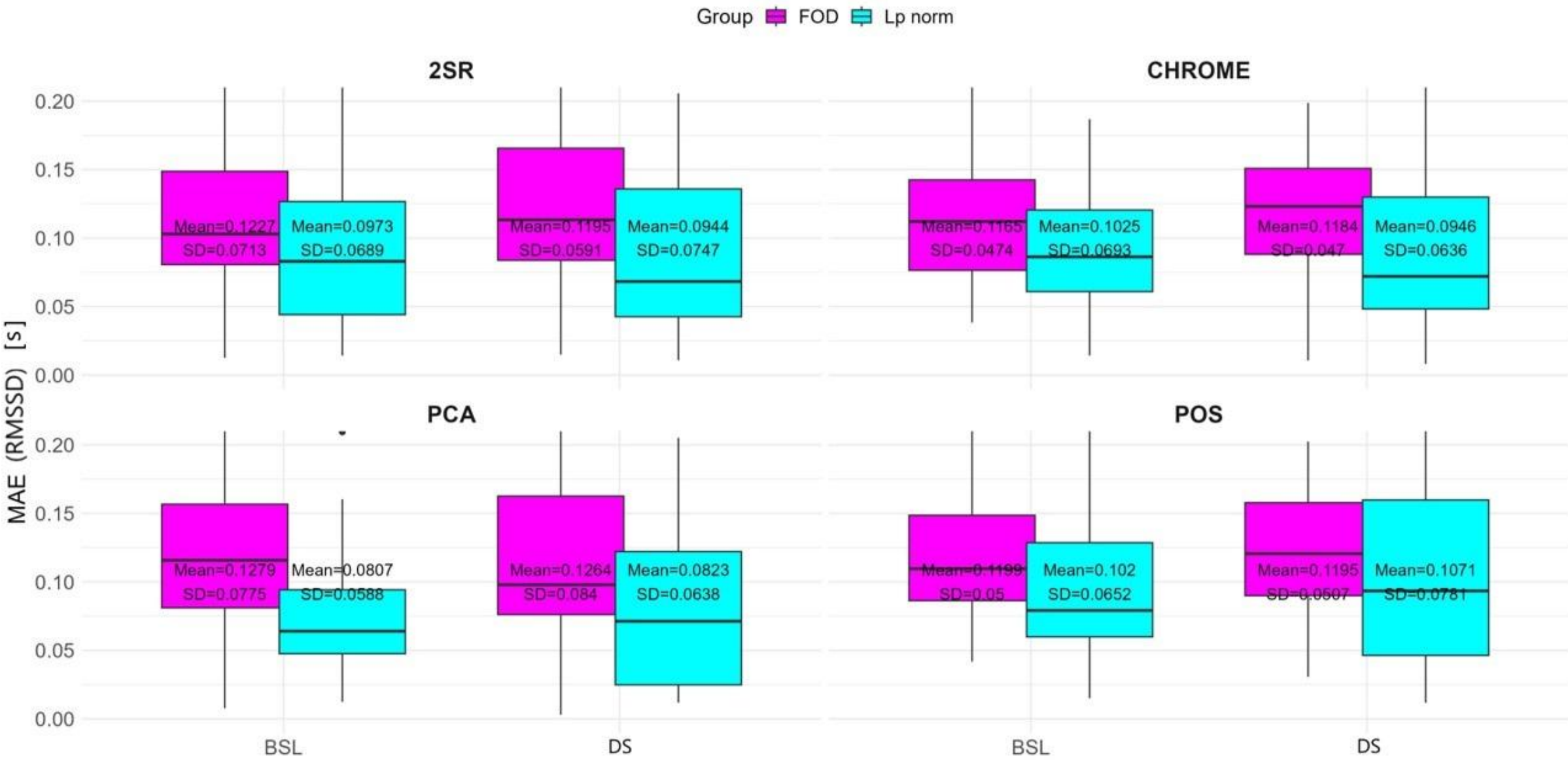

Figure 10: MAE between RMSSD from the reference ECG and extracted RMSSD for two peak enhancement approaches per each rPPG method. Shown graphic corresponds to initially determined 20 SP.

Table 2 presents the results of the statistical analysis performed on all reference HRV measures and the corresponding rPPG-based HRV parameters (PR/HR, SDNN, and RMSSD), comparing the BSL and DS measurement periods. A comparison of the results obtained by the proposed method with the results from the literature is shown in Tables 3 and 4.

Table 1: IBI metrics when it is considered according to the Relaxed definition of the IBI metrics. Both 10 SP and 20 SP regions are initially determined, two peak enhancement approaches and four rPPG methods. The most common *p* values of *Lp* norm are 6 and 7, while GL FOD is between 1 and 1.4. The highest evaluation metric across states (BSL and DS), number of SP regions, and rPPG methods is highlighted in bold.

| | Precision | | | | | | | |
|---|---|---|---|---|---|---|---|---|
| rPPG Methods | 10 regions | | | | 20 regions | | | |
| | *Lp* norm | | GL FOD | | *Lp* norm | | GL FOD | |
| | BSL | DS | BSL | DS | BSL | DS | BSL | DS |
| 2SR | 0.932 | 0.932 | 0.865 | 0.881 | 0.929 | 0.926 | 0.863 | 0.880 |
| CHROM | 0.925 | 0.918 | 0.859 | 0.870 | 0.924 | 0.920 | 0.859 | 0.873 |
| POS | 0.917 | 0.917 | 0.861 | 0.862 | 0.920 | 0.916 | 0.861 | 0.865 |
| PCA | **0.952** | 0.928 | 0.892 | 0.865 | 0.948 | 0.927 | 0.887 | 0.871 |
| Inverted PCA | 0.942 | 0.933 | 0.885 | 0.874 | 0.939 | 0.931 | 0.881 | 0.877 |
| | Recall | | | | | | | |
| rPPG Methods | 10 regions | | | | 20 regions | | | |
| | *Lp* norm | | GL FOD | | *Lp* norm | | GL FOD | |
| | BSL | DS | BSL | DS | BSL | DS | BSL | DS |
| 2SR | 0.905 | 0.883 | 0.900 | 0.872 | 0.906 | 0.883 | 0.895 | 0.868 |
| CHROM | 0.907 | 0.876 | 0.899 | 0.872 | 0.905 | 0.884 | 0.899 | 0.878 |
| POS | 0.899 | 0.867 | 0.889 | 0.861 | 0.905 | 0.879 | 0.895 | 0.873 |
| PCA | 0.910 | 0.842 | 0.896 | 0.835 | **0.911** | 0.865 | 0.883 | 0.814 |
| Inverted PCA | 0.898 | 0.851 | 0.892 | 0.842 | 0.902 | 0.865 | 0.891 | 0.841 |
| | F1 score | | | | | | | |
| rPPG Methods | 10 regions | | | | 20 regions | | | |
| | *Lp* norm | | GL FOD | | *Lp* norm | | GL FOD | |
| | BSL | DS | BSL | DS | BSL | DS | BSL | DS |
| 2SR | 0.918 | 0.907 | 0.882 | 0.877 | 0.918 | 0.904 | 0.879 | 0.874 |
| CHROM | 0.916 | 0.896 | 0.879 | 0.871 | 0.914 | 0.901 | 0.878 | 0.876 |
| POS | 0.908 | 0.891 | 0.874 | 0.862 | 0.912 | 0.897 | 0.877 | 0.869 |
| PCA | **0.931** | 0.883 | 0.894 | 0.850 | 0.929 | 0.895 | 0.885 | 0.842 |
| Inverted PCA | 0.919 | 0.890 | 0.889 | 0.858 | 0.920 | 0.897 | 0.886 | 0.858 |

Table 2: Results of statistical analysis between BSL and DS periods, for the reference HRV parameter from the ECG and extracted rPPG-based HRV parameters. The most common *p* values of *Lp* norm are 6 and 7, while GL FOD is 1-1.4.

| | PR/HR | | | | | | | |
|---|---|---|---|---|---|---|---|---|
| rPPG Methods | 10 regions | | | | 20 regions | | | |
| | *Lp* norm | | GL FOD | | *Lp* norm | | GL FOD | |
| | BSL | DS | BSL | DS | BSL | DS | BSL | DS |
| 2SR | $p$<0.01, $\delta$ = 0.21 | | $p$<0.01, $\delta$ = 0.21 | | $p$<0.01, $\delta$ = 0.23 | | $p$<0.01, $\delta$ = 0.23 | |
| CHROM | $p$<0.01, $\delta$ = 0.25 | | $p$<0.01, $\delta$ = 0.25 | | p<0.01, $\delta$ = 0.24 | | $p$<0.01, $\delta$ = 0.24 | |
| POS | $p$<0.01, $\delta$ = 0.24 | | $p$<0.01, $\delta$ = 0.21 | | $p$<0.01, $\delta$ = 0.24 | | $p$<0.01, $\delta$ = 0.23 | |
| PCA | $p$<0.01, $\delta$ = 0.18 | | $p$<0.01, $\delta$ = 0.22 | | $p$<0.01, $\delta$ = 0.22 | | $p$<0.01, $\delta$ = 0.27 | |
| Ref. ECG | $p$ <0.01, $\delta$ = 0.94 | | | | | | | |
| | SDNN | | | | | | | |
| rPPG Methods | 10 regions | | | | 20 regions | | | |
| | *Lp* norm | | GL FOD | | *Lp* norm | | GL FOD | |
| | BSL | DS | BSL | DS | BSL | DS | BSL | DS |
| 2SR | $p$ = 0.21, $\delta$ = 0.045 | | $p$ = 0.47, $\delta$ = 0.025 | | $p$ = 0.24, $\delta$ = 0.041 | | $p$ = 0.62, $\delta$ = 0.018 | |
| CHROM | $p$ = 0.70, $\delta$ = 0.014 | | $p$ = 0.40, $\delta$ = 0.030 | | $p$ = 0.16, $\delta$ = 0.049 | | $p$ = 0.82, $\delta$ = 0.008 | |
| POS | $p$ = 0.93, $\delta$ = 0.003 | | $p$ = 0.15, $\delta$ = 0.052 | | $p$ = 0.47, $\delta$ = 0.026 | | $p$ = 0.56, $\delta$ = 0.021 | |
| PCA | $p$ = 0.66, $\delta$ = 0.016 | | $p$ = 0.23, $\delta$ = 0.043 | | $p$ = 0.15, $\delta$ = 0.051 | | $p$ = 0.71, $\delta$ = 0.013 | |
| Ref. ECG | $p$ = 0.74, $\delta$ = 0.024 | | | | | | | |
| | RMSSD | | | | | | | |
| rPPG Methods | 10 regions | | | | 20 regions | | | |
| | *Lp* norm | | GL FOD | | *Lp* norm | | GL FOD | |
| | BSL | DS | BSL | DS | BSL | DS | BSL | DS |
| 2SR | $p$ = 0.25, $\delta$ = 0.040 | | $p$ = 0.34, $\delta$ = 0.034 | | $p$ = 0.14, $\delta$ = 0.052 | | $p$ = 0.52, $\delta$ = 0.029 | |
| CHROM | $p$ = 0.51, $\delta$ = 0.024 | | $p$ = 0.78, $\delta$ = 0.010 | | $p$ = 0.19, $\delta$ = 0.047 | | $p$ = 0.77, $\delta$ = 0.011 | |
| POS | $p$ = 0.93, $\delta$ = 0.003 | | $p$ = 0.40, $\delta$ = 0.030 | | $p$ = 0.39, $\delta$ = 0.031 | | $p$ = 0.67, $\delta$ = 0.015 | |
| PCA | $p$ = 0.43, $\delta$ = 0.028 | | $p$ = 0.17, $\delta$ = 0.049 | | $p$ = 0.14, $\delta$ = 0.052 | | $p$ = 0.94, $\delta$ = 0.003 | |
| Ref. ECG | $p$ = 0.37, $\delta$ = 0.064 | | | | | | | |

Table 3: A comparative analysis of the best results of our proposed methodology during BSL against results from literature also in conditions without expressed movements. Superior performance is denoted by the minimum average mean value achieved among the evaluated methods. The most common *p* values of *Lp* norm are 6 and 7, while GL FOD is between 1 and 1.4. The highest evaluation metric across different studies for each Method is highlighted in bold.

| Method | Environment | MAE (PR) [bpm] | RMSE (PR) [bpm] | MAE (SDNN) [s] | RMSE (SDNN) [s] | MAE (RMSSD) [s] | RMSE (RMSSD) [s] |
|---|---|---|---|---|---|---|---|
| The best achieved results | BSL | 2.22 ±2.26 | 2.91±2.69 | 0.061±0.040 | 0.087±0.046 | 0.081±0.059 | 0.119±0.079 |
| [55] POS | Still | **1.10** | / | 0.062 | / | 0.093 | / |
| [56] CHROM | Natural movements | 6.86 | 15.57 | 0.056 | **0.065** | 0.099 | **0.109** |
| [56] POS | Natural movements | 4.25 | 12.06 | 0.078 | 0.083 | 0.142 | 0.149 |
| [56] ICA | Natural movements | 6.05 | 13.30 | 0.077 | 0.082 | 0.136 | 0.145 |
| [57] ICA | "Static subject" | / | / | Absolute error range [**0.028** s – 0.074 s] | / | Absolute error range [0.065 s – 0.114 s] | / |
| [57] CHROM | "Static subject" | / | / | Absolute error range [0.034 s – 0.056 s] | / | Absolute error range [**0.056** s – 0.101 s] | / |
| [30] | Sitting still | / | / | 0.012 | / | 0.016 | / |
| [58] POS | Natural movements (dataset1) | / | / | 0.049±0.045 | / | 0.073±0.057 | / |
| [58] POS | Natural movements | / | / | 0.107±0.051 | / | 0.108±0.051 | / |
| [59] 3DCNN | Sitting, natural movements (dataset2) | 2.09 | 7.30 | / | / | / | / |
| [60] PCA | Still sitting | / | 11.08 | / | / | / | / |
| [60] CHROM | Still sitting | / | **2.39** | / | / | / | / |
| [60] POS | Still sitting | / | 3.69 | / | / | / | / |
| [61] POS | Still, hospital environment | 2.30 | 3.20 | / | / | / | / |
| | | | | IBI, Metrocs | | | |
| | | | | Accuracy | Precision | Recall | F1 score |
| The best achieved results in our approach | | | | / | **0.95** | 0.91 | 0.93 |
| [62] CHROM | Rest, without movements | 4.17 | 7.70 | 0.88 | / | / | / |
| [62] POS | Rest, without movements | 2.10 | 5.39 | 0.94 | / | / | / |
| [63] PPG *vs* ECG | Rest, without movements | / | / | Sensitivity | / | / | 0.94 |

Table 4: A comparative analysis of the best results of our proposed methodology during DS against results from literature also in conditions with highly expressed movements. Superior performance is denoted by the minimum average mean value achieved among the evaluated methods. The most common *p* values of *Lp* norm are 6 and 7, while GL FOD is between 1 and 1.4. The highest evaluation metric across different studies for each Method is highlighted in bold.

| Method | Environment | MAE (PR) [bpm] | RMSE (PR) [bpm] | MAE (SDNN) [s] | RMSE (SDNN) [s] | MAE (RMSSD) [s] | RMSE (RMSSD) [s] |
|---|---|---|---|---|---|---|---|
| The best achieved results | DS | **1.92±1.72** | 2.45±2.00 | **0.071±0.044** | 0.097±0.055 | **0.082±0.064** | 0.111±0.082 |
| [55] POS | Interactive tasks | 3.08 | / | 0.088 | / | 0.120 | / |
| [57] ICA | "Frequent motion" | / | / | Absolute error range [0.136 s – 0.189 s] | / | Absolute error range [0.164 s – 0.219 s] | / |
| [57] CHROM | "Frequent motion" | / | / | Absolute error range [0.075 s – 0.076 s] | / | Absolute error range [0.096 s – 0.137 s] | / |
| [64] 2SR | Wide range of activity | 3.72 | / | / | / | / | / |
| [64] POS | Wide range of activity | 2.07 | / | / | / | / | / |

# 4. Discussion

Concerning data quality, the most prevalent cause of segment rejection across all methods is face detection failure, with typical exclusion is 6-10 recordings out of approximately 70 segments. This type of failure is inherently linked to the video acquisition conditions, particularly head pose variability, as evidenced by an extreme case in which 68 out of 74 segments are discarded due to the participant maintaining a lowered head position throughout the recording. Such cases highlight the sensitivity of facial landmark-based pipelines to participant compliance and recording setup and underline the importance of monitoring head position during data acquisition.

Beyond face detection failures, segment rejection based on the SNR threshold (-17 dB) [41] reveals notable differences between the evaluated rPPG methods. The PCA-based approach exhibited the highest rejection rate, with approximately 20 segments discarded per recording, while other three methods (2SR, CHROM, and POS) showed greater robustness, with up to 5 rejections per recording. This discrepancy can be attributed to the fundamental nature of PCA [65]. Unlike methods that exploit predefined assumptions about the spectral or chromatic properties of the rPPG signal, PCA identifies components purely based on variance structure [65]. As a result, it is inherently susceptible to amplifying noise that happens to carry high variance. In conditions where the SNR of the raw rPPG signal is low, as is often the case in DS recording environments [40], PCA may yield a dominant component that does not correspond to cardiac activity, leading to segments being flagged and excluded by the SNR-based quality criterion. This behavior, while limiting in terms of data yield, is consistent with findings reported in the broader rPPG literature, which have similarly noted the sensitivity of PCA to noise when applied without additional constraints [65].

## 4.1. Optimal Facial Regions and Enhancement Parameters

Our findings show that the most informative rPPG regions are the cheeks, or cheeks combined with the forehead, consistent with previous studies [11, 23, 59, 66]. Cheeks are selected most frequently, with a strong left-side preference likely due to lighting, while right-sided regions appear rarely. Region preferences are largely consistent across rPPG methods, indicating they are driven by intrinsic skin perfusion rather than the algorithm used. For signal processing, optimal *Lp* norm values clustered around $p = 6–7$, sharpening peaks via stronger suppression of low-amplitude components, though very high $p$ may distort beat variability. In contrast, GL FOD performed the best with lower orders (1.1–1.4), balancing peak enhancement and noise suppression. Detailed Discussion on this topic is presented in [35].

## 4.2. Signal Quality Assessment

The signal quality of the extracted rPPG signals is evaluated using three SNR definitions, presented in Figs. 6, A9, and A10 (in Supplementary Materials [35]), all corresponding to the 20 SP configuration: a narrow band of ±0.1 Hz centered at the reference ECG frequency (Fig. A9), a broad band spanning 0.7 Hz–3 Hz (Fig. 6), and a narrow band of ±0.1 Hz centered at the dominant rPPG frequency (Fig. A10). Across all three figures (Figs. 6, A9, and A10), SNR values are consistently lower during DS compared to BSL (Table A5), which is attributable to increased head motion or potential illumination changes [40] introduced by the driving simulator.

A further consistent observation is the close agreement between SNR values obtained with the *Lp* norm and GL FOD approaches for any given rPPG method. This suggests that the peak enhancement strategy does not meaningfully influence which facial regions are identified as informative. Regions carrying a strong cardiac-frequency component are selected regardless of the enhancement method applied, indicating that region utility is an intrinsic property of the signal rather than an artifact of the downstream processing choice. Extended discussion on this topic is presented in Supplementary Materials [35].

## 4.3. Heart Rate Variability Estimation Success

Bland–Altman analysis (Fig. 7) shows strong agreement between rPPG-derived PR and ECG reference HR across all methods, with most errors for clinical use [67] within the ±5 bpm. Despite expected challenges during the DS task, error increases are modest, probably thanks to SNR-based segment rejection. The best performance is achieved with 2SR using 20 SP and GL FOD peak enhancement, yielding MAE/RMSE of 1.92±1.72 / 2.45±2.00 bpm under DS—substantially better than many literature results under both static and motion conditions [55-56, 60]. Using 20 SP regions consistently outperformed 10 SP, improving MAE and RMSE by ~30–37% for PR at the cost of doubled computation time. GL FOD generally produced lower errors than *Lp* norm ($p \approx 6–7$), especially under DS, while PCA showed higher rejection rates but competitive results when sufficient clean segments remained. HRV estimation (SDNN and RMSSD) proved more challenging due to sensitivity to IBI errors, yet the framework still achieved competitive or superior accuracy (the best SDNN MAE/RMSE: 0.061/0.087 s; RMSSD: 0.081/0.111 s) compared to literature values from less demanding conditions [55, 57-58]. Differences between BSL and DS in our study are consistently small for both PR and HRV parameters, highlighting the effectiveness of adaptive region selection, SNR filtering, and peak enhancement in mitigating motion artifacts.

## 4.4. Performance of Detection of Inter-Beat-Intervals

IBI detection performance is evaluated using Precision, Recall, and F1 score under three matching tolerances: Strict, Medium, and Relaxed. As expected, F1 scores increased markedly with tolerance, ranging from 0.37–0.43 (Strict) to 0.49–0.58 (Medium), and reaching 0.85–0.93 under the Relaxed criterion. These results are competitive with literature values [62-63], despite using stricter temporal tolerances (250–350 ms) than prior studies. Across all criteria, 2SR and CHROM consistently achieved slightly higher F1 scores than POS and PCA. Differences between BSL and DS conditions remained modest, underscoring the effectiveness of SNR-based segment rejection. While no clear advantage between peak enhancement methods appeared under Strict and Medium criteria, the *Lp* norm showed a slight but consistent edge over GL FOD under the Relaxed criterion, likely due to GL FOD introducing variable temporal shifts in peak locations.

## 4.5. Statistical Analysis of HRV Parameters During Baseline and Driving Simulation Conditions

For the PR/HR parameter, the reference ECG data yields a statistically significant difference between BSL and DS conditions ($p < 0.01$, $\delta = 0.94$) (Table 2), confirming that HR/PR is meaningfully elevated during the DS relative to BSL, as would be expected given the cognitive and physiological demands of the task [68]. Importantly, this significant difference is consistently reproduced across all rPPG methods, SP configurations, and peak enhancement approaches, with $p < 0.01$ in all cases. The effect sizes obtained from the rPPG-derived PR values ($\delta$ ranging approximately from 0.18 to 0.27) are lower than the reference ECG effect size, but still notable, which is attributable to the greater measurement noise inherent to contactless rPPG-based estimation attenuating the magnitude of the observed effect. This suggests that during the low-motion BSL condition, the choice of processing configuration has little influence on the ability to detect PR differences, whereas under DS conditions the specific method becomes more consequential, as increased motion and physiological arousal expose differences in robustness across the evaluated approaches.

For SDNN and RMSSD, the reference ECG did not reveal a statistically significant difference between BSL and DS conditions ($p = 0.74$ and $p = 0.37$, respectively), with small corresponding effect sizes ($\delta = 0.024$ and $\delta = 0.064$). This pattern is uniformly mirrored in the rPPG-derived results: across all methods, configurations, and peak enhancement approaches, none of the extracted SDNN or RMSSD values yield a statistically significant BSL *vs* DS difference. This concordance indicates that the proposed pipeline not only provides acceptable point estimates of HRV parameters, but also preserves the statistical structure of the underlying physiological data sufficiently to support analyses. The consistency of this behaviour across all evaluated rPPG methods and both SP configurations further underscores the robustness of the framework as a tool for contactless autonomic monitoring in the DS environment.

## 4.6. Limitations and Future Directions

Despite its contributions presented in [35], this study has several limitations that warrant consideration and suggest avenues for future investigation:

1. Combination of informative sub-regions (left or right cheek with the forehead) to obtain a more comprehensive result as suggested in [69] can be performed. Namely, future work should investigate whether combining multiple highly informative SP-based facial sub-regions, rather than relying on a single region, can further improve the robustness and accuracy of rPPG estimation, especially under challenging illumination conditions. Specifically, we could be exploring a tilling and aggregation technique as proposed in [70] or by exploring similarity evaluation metrics for sub-region selection [71] and proper weightening combination as performed in [69] or simple averaging as proposed in [72].

2. As a potential improvement, in cases where face detection fails, specific facial regions (*e.g.*, eyes, nose, or mouth) could be detected and used to infer the approximate face location and define the region of interest for rPPG signal extraction.

3. The proper camera positioning in relation to the lighting sources for the optimal or sub-optimal extraction of an accurate and robust rPPG-based HRV parameters [25]. Such an analysis can be performed to analyze potential performance differences between facial regions located on the left and right sides of the face. It would be of the most importance to quantify the impact of asymmetric illumination and shadowing. Also, some methodological constraints should be taken into account - as scene brightness in the simulation itself, especially during simulated or real-life night driving conditions [73-74].

4. Detection of pathological beats (Fig. 11) - additional research direction could focus on extending the proposed framework toward the detection of pathological or abnormal heartbeats (approved by cardiologist), enabling a more comprehensive assessment of cardiovascular dynamics. The pathological beat is reflected in the rPPG waveform as an anomalous deflection which, unlike in ECG, cannot be distinguished from a true cardiac peak and therefore acts as noise in the IBI detection process. The designed processing framework could not detect pathological changes and even introduced larger errors in IBIs detection. An improved and dedicated processing framework may pave the way for rPPG application in clinical environments. Although such an approach is rather novel, there are previously made attempts to assess rPPG in patients and even to detect cardiac ejections [75].
5. Variable window sizes are not explored in detail in this study as we rely on the previous recommendation to use 20 s long segments [23, 59, 66] as the first step in the processing workflow (Fig. 1). This choice represents a deliberate compromise: a minimum of approximately 10 s is generally considered necessary for reliable PR estimation [76-77], while SDNN and RMSSD require a sufficient number of IBI samples to yield stable and meaningful estimates, precluding the use of very short windows. At the other extreme, our pilot analysis indicated that segments of 30 s or longer introduce increased head pose drift and less stable SP localization, consistent with the observations reported by Bobbia et al. [26], which can degrade the spatial correspondence of the defined ROIs across frames and consequently reduce rPPG signal quality.
6. Skin color variations are not considered since all participants are Caucasian. However, this factor is relevant, because the proposed method is sensitive to skin color [78], and cameras are typically optimized to capture lighter skin tones [15, 25].
7. The dataset used in this study is limited in size and demographic diversity. Future work could explore the use of synthetic data generation techniques — such as computer graphics-based video simulation or generative neural models — to augment datasets with a broader range of skin types, physiological states, and motion conditions, although a known sim-to-real performance gap remains a challenge for such approaches. [25]
8. Signal polarity in the present study is addressed through a fixed inversion strategy based on a pilot phase analysis. However, as demonstrated in the Supplementary Materials [35], the phase differences between the raw R, G, and B channels vary across recordings [38], and the application of different rPPG methods (2SR, CHROM, POS, and PCA) further modifies these phase relationships in a method-dependent manner, potentially resulting in phase offsets that deviate from a simple inversion (180° shift). An adaptive phase correction mechanism, estimating the alignment on a per-segment or per-subject basis, could therefore improve IBI detection more effectively than a fixed inversion. The benefit of even a simple correction is illustrated by the PCA results in the Supplementary Materials [35], where inversion of the PCA-derived signal improved Precision under the Medium criterion from 0.543 to 0.575 for the GL FOD 20 SP configuration (Table A7 in [35]), suggesting that a phase alignment strategy may be a useful direction for future work.

# 5. Conclusion

This study presented a contactless rPPG-based pipeline for the estimation of pulse rate and HRV parameters from facial video in a driving simulator environment. The proposed framework integrates SP-based adaptive facial region selection with two peak enhancement approaches — the *Lp* norm and the GL FOD— evaluated in combination with four rPPG methods (2SR, CHROM, POS, and PCA) using a comprehensive set of metrics encompassing PR, SDNN, and RMSSD estimation accuracy, SNR, and IBI detection performance.

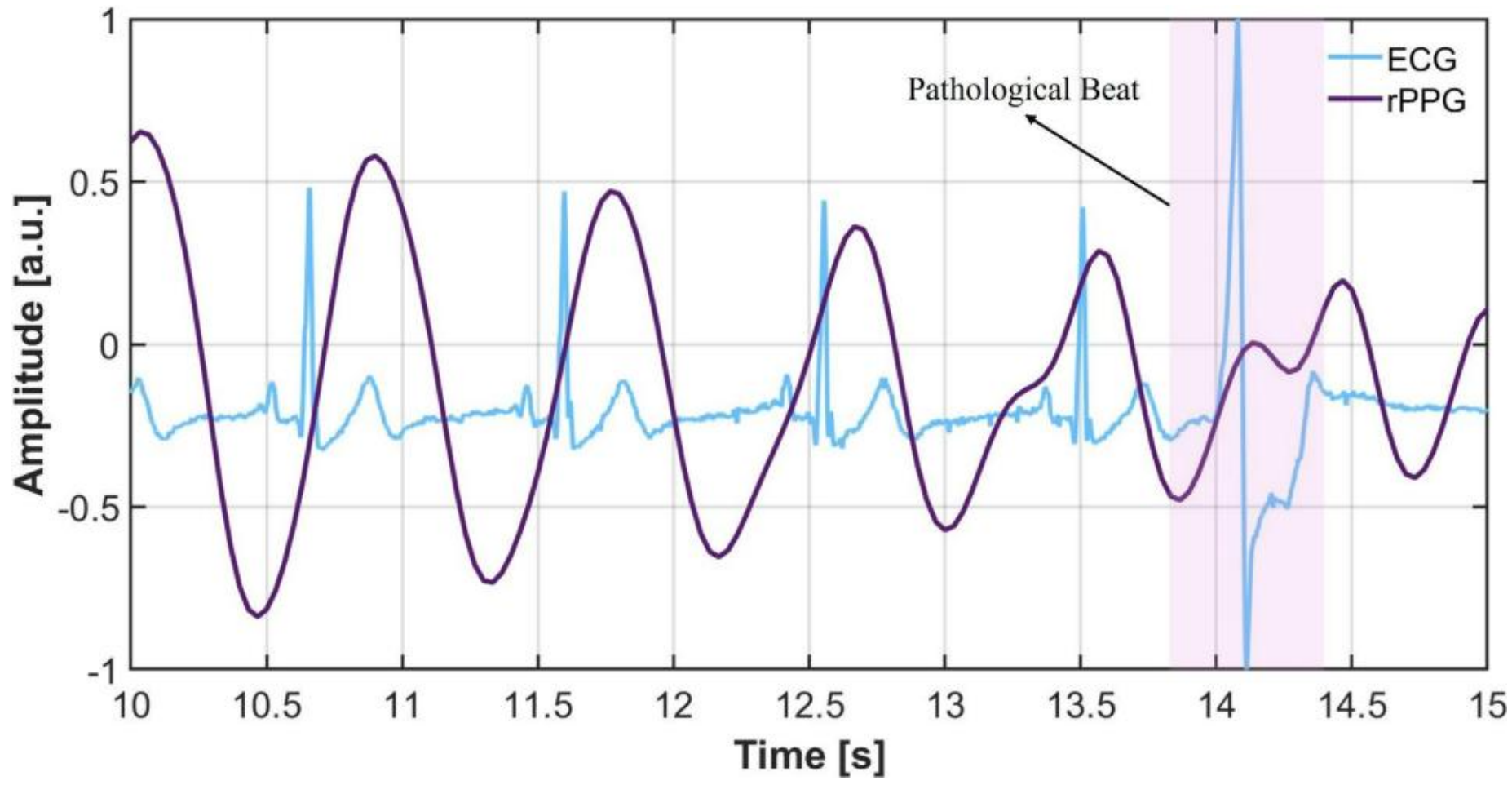

Figure 11: Example of a 20 s long ECG segment containing a pathological beat (confirmed by a cardiologist) and the corresponding rPPG signal extracted using the CHROM method with *Lp* norm ($p = 7$) during BSL. Please, note that the proposed method is not sensitive enough to detect pathological beat.

PR estimation achieved clinically acceptable accuracy across virtually all configurations, with mean errors remaining below the 5 bpm threshold associated with FDA-approved devices [67]. SDNN and RMSSD results, while more challenging due to the sensitivity of HRV metrics to peak localization errors, are consistent across methods and conditions, with the statistical structure of the ECG reference data faithfully reproduced by all rPPG-derived parameters. Differences in MAE and RMSE between BSL and DS remained modest throughout, reflecting the effectiveness of the SNR-based segment rejection in limiting the impact of signal degradation under more demanding recording conditions. These findings demonstrate the feasibility of fully contactless autonomic monitoring in dynamic environments, extending rPPG technology beyond controlled laboratory settings and baseline measurements. By enabling robust multi-feature HRV estimation from facial videos, the proposed framework offers a viable non-invasive alternative to traditional physiological sensing in driving research, laying the groundwork for future real-world deployment of camera-based driver monitoring systems oriented toward road safety and adaptive human-machine interaction.

Based on the overall results, 2SR with GL FOD (fractional order of 1.0–1.4, 20 SP) is recommended for PR estimation, while CHROM with *Lp* norm ($p = 6$–$7$, 20 SP) is preferred for SDNN and RMSSD estimation. More generally, when precise IBI estimation is the primary objective, the *Lp* norm is recommended over GL FOD, as the subtle and low-amplitude nature of rPPG signal variations makes them more susceptible to the peak displacement introduced by fractional-order differentiation (an effect less problematic in ECG where sharp waveform morphology makes derivative-based enhancement particularly effective) potentially compromising beat-to-beat temporal accuracy.

**Acknowledgement:** We wish to thank Professor Jaka Sodnik and Assistant Professor Sara Stančin from the Faculty of Electrical Engineering, University of Ljubljana (FE), as well as student Filipa Zoa (from the FE and the Faculty of Engineering, University of Porto) for their valuable discussions and contributions during initial stages of this research. Also, we would like to thank Professor Tomislav B. Šekara from the University of Belgrade - School of Electrical Engineering for his valuable and constructive feedback regarding the use of the fractional derivative and the *Lp* norm, which contributed to this work. We extend our sincere thanks to Jelena Medarević, PhD student at the Faculty of Electrical Engineering, University of Ljubljana and to Smilja Stokanović, PhD student from the University of Belgrade - School of Electrical Engineering for their valuable assistance for protocol design and for conducting the measurements. Finally, we would like to thank Nikola Radovanović, MD PhD from the Pacemaker Center, University Clinical Center of Serbia and Faculty of Medicine, University of Belgrade for his valuable insights into possibilities for rPPG-based detection of extra systoles.

**Funding:** Đorđe D. Nešković visited the Faculty of Electrical Engineering, University of Ljubljana as part of a collaborative research initiative in March 2025 within the framework of the European Union Horizon Europe research and innovation programme under the FRODDO (Federated cybeR-physical infrastructure for ODD cOntinuity) project (Grant Agreement No. 101147819) - the visit was related to research presented in this study. Nadica Miljković acknowledges financial support from Grant No. 451-03-34/2026-03/200103, funded by the Ministry of Science, Technological Development, and Innovation of the Republic of Serbia.

**Note on Generative Use of Artificial Intelligence**: The Authors utilized GPT-5-mini (ChatGPT) to enhance clarity and language and the Authors are fully responsible for the publication content after using this tool.

**CRediT Author Statement**: Đorđe D. Nešković: Visualization, Software, Methodology, Formal analysis, Investigation, Writing - original draft, Nadica Miljković: Validation, Methodology, Investigation, Conceptualization, Writing - review & editing.

# References

1. Cao, R., Azimi, I., Sarhaddi, F., Niela-Vilen, H., Axelin, A., Liljeberg, P., & Rahmani, A. M. (2022). Accuracy assessment of oura ring nocturnal heart rate and heart rate variability in comparison with electrocardiography in time and frequency domains: comprehensive analysis. *Journal of Medical Internet Research*, *24*(1), e27487. https://doi.org/10.2196/27487 .
2. Dudarev, V.; Barral, O.; Zhang, C.; Davis, G.; Enns, J.T. On the reliability of wearable technology: A tutorial on measuring heart rate and heart rate variability in the wild. Sensors 2023, 23, 5863. https://doi.org/10.3390/s23135863.
3. Barka, R. E., & Politis, I. (2024). Driving into the future: A scoping review of smartwatch use for real-time driver monitoring. *Transportation Research Interdisciplinary Perspectives*, *25*, 101098. https://doi.org/10.1016/j.trip.2024.101098.
4. Prucnal, M. A., Polak, A. G., & Kazienko, P. (2025). Improving the quality of pulse rate variability derived from wearable devices using adaptive, spectrum and nonlinear filtering. *Biomedical Signal Processing and Control*, *102*, 107336. https://doi.org/10.1016/j.bspc.2024.107336.
5. Gaur, P., Temple, D. S., Hegarty-Craver, M., Boyce, M. D., Holt, J. R., Wenger, M. F., ... & Dausch, D. E. (2024). Continuous monitoring of heart rate variability in free-living conditions using wearable sensors: Exploratory observational study. *JMIR Formative Research*, *8*, e53977. https://doi.org/10.2196/53977.
6. Brookhuis, K. A., & De Waard, D. (2010). Monitoring drivers' mental workload in driving simulators using physiological measures. *Accident Analysis & Prevention*, 42(3), 898-903. https://doi.org/10.1016/j.aap.2009.06.001.
7. Xiao, H., Liu, T., Sun, Y., Li, Y., Zhao, S., & Avolio, A. (2024). Remote photoplethysmography for heart rate measurement: A review. *Biomedical Signal Processing and Control*, *88*, 105608. https://doi.org/10.1016/j.bspc.2023.105608.
8. Sakib, S., Hasan, Z., & Roy, N. (2025). A state-of-the-art survey of remote photoplethysmography for contactless health parameters sensing. *Wiley Interdisciplinary Reviews: Data Mining and Knowledge Discovery*, *15*(3), e70039. https://doi.org/10.1002/widm.70039.
9. Seo, H., Kim, S., & Lee, E. C. (2025). Estimation of respiratory signals from remote photoplethysmography of RGB facial videos. *Electronics*, *14*(11), 2152. https://doi.org/10.3390/electronics14112152.
10. Lewandowska, M.; Nowak, J. Measuring pulse rate with a webcam. *J. Med. Imaging Health Inform*. 2012, 2, 87–92. https://doi.org/10.1166/jmihi.2012.1064.
11. Kwon, S.; Kim, H.; Park, K.S. Validation of heart rate extraction using video imaging on a built-in camera system of a smartphone. *In Proceedings of the 2012 Annual International Conference of the IEEE Engineering in Medicine and Biology Society*, San Diego, CA, USA, 28 August–1 September 2012; IEEE: Piscataway, NJ, USA, 2012; pp. 2174–2177. https://doi.org/10.1109/EMBC.2012.6346392.
12. Poh, M.Z.; McDuff, D.J.; Picard, R.W. Non-contact, automated cardiac pulse measurements using video imaging and blind source separation. *Opt. Express* 2010, 18, 10762–10774. https://doi.org/10.1364/OE.18.010762.
13. Ernst, H.; Scherpf, M.; Malberg, H.; Schmidt, M. Optimal color channel combination across skin tones for remote heart rate measurement in camera-based photoplethysmography. *Biomedical Signal Processing and Control* 2021, 68, 102644. https://doi.org/10.1016/j.bspc.2021.102644.

14. Renner, P.; Gleichauf, J.; Winkelmann, S. Non-Contact In-Car Monitoring of Heart Rate: Evaluating the Eulerian Video Magnification Algorithm in a Driving Simulator Study. *In Proceedings of the Mensch und Computer 2024*, Karlsruhe, Germany, 1–4 September 2024; pp. 651–654. https://doi.org/10.1145/3670653.3677493.
15. Dasari, A., Prakash, S. K. A., Jeni, L. A., & Tucker, C. S. (2021). Evaluation of biases in remote photoplethysmography methods. *NPJ digital medicine*, *4*(1), 91. https://doi.org/10.1038/s41746-021-00462-z.
16. Wang, W., Den Brinker, A. C., Stuijk, S., & De Haan, G. (2016). Algorithmic principles of remote PPG. *IEEE Transactions on Biomedical Engineering*, 64(7), 1479-1491. https://doi.org/10.1109/TBME.2016.2609282.
17. Elgendi, M., Martinelli, I., & Menon, C. (2024). Optimal signal quality index for remote photoplethysmogram sensing. *npj Biosensing*, 1, 5. https://doi.org/10.1038/s44328-024-00002-1.
18. De Haan, G., & Jeanne, V. (2013). Robust pulse rate from chrominance-based rPPG. *IEEE Transactions on Biomedical Engineering*, 60(10), 2878-2886. https://doi.org/10.1109/TBME.2013.2266196.
19. Wang, W., Stuijk, S., & De Haan, G. (2015). A novel algorithm for remote photoplethysmography: Spatial subspace rotation. *IEEE Transactions on Biomedical Engineering*, 63(9), 1974-1984. https://doi.org/10.1109/TBME.2015.2508602.
20. Wu, H.Y.; Rubinstein, M.; Shih, E.; Guttag, J.; Durand, F.; Freeman, W. Eulerian video magnification for revealing subtle changes in the world. ACM Trans. Graph. (TOG) 2012, 31, 1–8. https://doi.org/10.1145/2185520.2185561.
21. Miljković, N.; Trifunovi´c, D. Pulse rate assessment: Eulerian video magnification vs. electrocardiography recordings. *In Proceedings of the 12th Symposium on Neural Network Applications in Electrical Engineering (NEUREL),* Belgrade, Serbia, 25–27 November 2014; IEEE: Piscataway, NJ, USA, 2012; pp. 17–20. https://doi.org/10.1109/NEUREL.2014.7011447 .
22. Nešković, Đ. D., Stojmenova Pečečnik, K., Sodnik, J., & Miljković, N. (2025). Contactless pulse rate assessment: Results and insights for application in driving simulators. *Applied Sciences*, *15*(17), 9512. https://doi.org/10.3390/app15179512.
23. Nagar, S., Hasegawa-Johnson, M., Beiser, D. G., & Ahuja, N. (2024). R2I-rPPG: A robust region of interest selection method for remote photoplethysmography to extract heart rate. *arXiv preprint arXiv:2410.15851*. https://doi.org/10.48550/arXiv.2410.15851.
24. Pai, A., Veeraraghavan, A., & Sabharwal, A. (2021). HRVCam: robust camera-based measurement of heart rate variability. *Journal of biomedical optics*, *26*(2), 022707-022707. https://doi.org/10.1117/1.JBO.26.2.022707.
25. McDuff, D. (2023). Camera measurement of physiological vital signs. *ACM Computing Surveys*, 55(9), 1-40. https://doi.org/10.1145/3558518.
26. Bobbia, S., Luguern, D., Benezeth, Y., Nakamura, K., Gomez, R., & Dubois, J. (2018). Real-time temporal superpixels for unsupervised remote photoplethysmography. *In Proceedings of the IEEE Conference on Computer Vision and Pattern Recognition Workshops*, pp. 1341-1348.
27. Medarević, J., Miljković, N., Stojmenova Pečečnik, K., & Sodnik, J. (2025). Distress detection in VR environment using Empatica E4 wristband and Bittium Faros 360. *Frontiers in Physiology*, *16*, 1480018. https://doi.org/10.3389/fphys.2025.1480018.
28. Yu, S. G., Kim, S. E., Kim, N. H., Suh, K. H., & Lee, E. C. (2021). Pulse rate variability analysis using remote photoplethysmography signals. *Sensors*, *21*(18), 6241. https://doi.org/10.3390/s21186241.
29. Shaffer F., Ginsberg J. P. (2017). An overview of heart rate variability metrics and norms. *Frontiers in Public Health* 5, 258. https://doi.org/10.3389/fpubh.2017.00258.
30. Finžgar, M., & Podržaj, P. (2020). Feasibility of assessing ultra-short-term pulse rate variability from video recordings. *PeerJ*, *8*, e8342. https://doi.org/10.7717/peerj.8342.
31. Pander, T., Przybyła, T., & Czabański, R. (2013). An application of the Lp-norm in robust weighted averaging of biomedical signals. *Journal of Medical Informatics & Technologies*, *22.*
32. Tanasković, I., & Miljković, N. (2023). A new algorithm for fetal heart rate detection: Fractional order calculus approach. *Medical Engineering & Physics*, *118*, 104007. https://doi.org/10.1016/j.medengphy.2023.104007.
33. Miljković, N., Popović, N., Djordjević, O., Konstantinović, L., & Šekara, T. B. (2017). ECG artifact cancellation in surface EMG signals by fractional order calculus application. *Computer Methods and Programs in Biomedicine*, *140*, 259-264. https://doi.org/10.1016/j.cmpb.2016.12.017.
34. Stojmenova Pečečnik, K., Tomažič, S., Sodnik, J., (2023). Design of head-up display interfaces for automated vehicles. *International Journal of Human-Computer Studies.* 177, 103060. https://doi.org/10.1016/j.ijhcs.2023.103060.
35. Nešković D. Đ, Miljković N, (2026). Supplementary Material for Manuscript A Signal Extraction Approach for Remote Heart Rate Variability Assessment Using Proxy Measure in a Driving Simulator, Zenodo https://doi.org/10.5281/zenodo.20354210.
36. Li, P., Benezeth, Y., Nakamura, K., Gomez, R., & Yang, F. (2019). Model-based region of interest segmentation for remote photoplethysmography. In the *14th International Conference on Computer Vision Theory and Applications*, pp. 383-388, SCITEPRESS-Science and Technology Publications. https://dx.doi.org/10.5220/0007389803830388.
37. Benezeth, Y., Bobbia, S., Nakamura, K., Gomez, R., & Dubois, J. (2019). Probabilistic signal quality metric for reduced complexity unsupervised remote photoplethysmography. In *2019 13th International Symposium on Medical Information and Communication Technology (ISMICT)*, pp. 1-5, IEEE. https://doi.org/10.1109/ISMICT.2019.8744004.

38. Huang, Y., Huang, D., Huang, J., Lu, H., He, M., & Wang, W. (2023, July). Camera wavelength selection for multi-wavelength pulse transit time based blood pressure monitoring. In *2023 45th Annual International Conference of the IEEE Engineering in Medicine & Biology Society (EMBC)* (pp. 1-5). IEEE. https://doi.org/10.1109/EMBC40787.2023.10340068.
39. Saritas, T., Greber, R., Venema, B., Puelles, V. G., Ernst, S., Blazek, V., ... & Schlieper, G. (2019). Non-invasive evaluation of coronary heart disease in patients with chronic kidney disease using photoplethysmography. *Clinical kidney journal*, *12*(4), 538-545. https://doi.org/10.1093/ckj/sfy135.
40. Xi, L., Wu, X., Chen, W., Wang, J., & Zhao, C. (2022). Weighted combination and singular spectrum analysis based remote photoplethysmography pulse extraction in low-light environments. *Medical engineering & physics*, *105*, 103822. https://doi.org/10.1016/j.medengphy.2022.103822.
41. Lin, Y. C., & Lin, Y. H. (2017). A study of color illumination effect on the SNR of rPPG signals. In *2017 39th Annual International Conference of the IEEE Engineering in Medicine and Biology Society (EMBC)*, pp. 4301-4304, IEEE. https://doi.org/10.1109/EMBC.2017.8037807.
42. Pan, J., & Tompkins, W. J. (1985). A real-time QRS detection algorithm. *IEEE Tansactions on Biomedical Engineering*, (3), 230-236. https://doi.org/10.1109/TBME.1985.325532.
43. Guler, S., Golparvar, A., Ozturk, O., Dogan, H., & Yapici, M. K. (2023). Optimal digital filter selection for remote photoplethysmography (rPPG) signal conditioning. *Biomedical Physics & Engineering Express*, 9(2), 027001. https://iopscience.iop.org/article/10.1088/2057-1976/acaf8a/meta.
44. Liao, G., Lu, H., Shan, C., & Wang, W. (2023). Plethysmographic waveform features and hemodynamic features for camera-based blood pressure estimation. *IEEE Transactions on Instrumentation and Measurement*, *73*, 1-14. https://doi.org/10.1109/TIM.2023.3338714.
45. Kantrowitz, A. B., Ben-David, K., Morris, M., Wittels, H. L., Wishon, M. J., McDonald, S. M., ... & Wittels, S. H. (2025). Pulse rate variability is not the same as heart rate variability: Findings from a large, diverse clinical population study. *Frontiers in Physiology*, 16, 1630032. https://doi.org/10.3389/fphys.2025.1630032.
46. Schäfer, A., & Vagedes, J. (2013). How accurate is pulse rate variability as an estimate of heart rate variability?: A review on studies comparing photoplethysmographic technology with an electrocardiogram. *International Journal of Cardiology*, 166(1), 15-29. https://doi.org/10.1016/j.ijcard.2012.03.119.
47. Allen, J., & Murray, A. (2002). Age-related changes in peripheral pulse timing characteristics at the ears, fingers and toes. *Journal of Human Hypertension*, *16*(10), 711-717. https://doi.org/10.1038/sj.jhh.1001478.
48. Allen, J., & Murray, A. (2000). Variability of photoplethysmography peripheral pulse measurements at the ears, thumbs and toes. *IEE Proceedings-Science, Measurement and Technology*, *147*(6), 403-407. https://doi.org/10.1049/ip-smt:20000846.
49. Hsieh, F., & Chen, T. L. (2025). A novel R-peak detection algorithm. *IEEE Access*, 13, 210351-210359. https://doi.org/10.1109/ACCESS.2025.3643153.
50. AAMI, A., & EC57, A. A. M. I. (2008). Testing and reporting performance results of cardiac rhythm and st segment measurement algorithms. *American National Standards Institute*, Arlington, VA, USA, 43.
51. Kamshilin, A. A., Sidorov, I. S., Babayan, L., Volynsky, M. A., Giniatullin, R., & Mamontov, O. V. (2016). Accurate measurement of the pulse wave delay with imaging photoplethysmography. *Biomedical optics express*, *7*(12), 5138-5147. https://doi.org/10.1364/BOE.7.005138.
52. Fariha, M. A. Z., Ikeura, R., Hayakawa, S., & Tsutsumi, S. (2020, June). Analysis of Pan-Tompkins algorithm performance with noisy ECG signals. In *Journal of Physics: Conference Series* (Vol. 1532, No. 1, p. 012022). IOP Publishing. https://iopscience.iop.org/article/10.1088/1742-6596/1532/1/012022/meta.
53. Mager, J. (2015). Automatic threshold selection of the peaks over threshold method. *Technical University of Munich Department of Mathematics,* 114, Master Thesis, Available online: https://mediatum.ub.tum.de/doc/1254349/document.pdf, Accessed on February 26, 2026.
54. Hess, M.R.; Kromrey, J.D. Robust confidence intervals for effect sizes: A comparative study of Cohen's d and Cliff's delta under non-normality and heterogeneous variances. *In Proceedings of the Annual Meeting of the American Educational Research Association*, San Diego, CA, USA, 12–16 April 2004; Volume 1.
55. Talukdar, D., De Deus, L. F., & Sehgal, N. (2023). The Evaluation of Remote Monitoring Technology Across Participants With Different Skin Tones. *Cureus*, *15*(9). https://doi.org/10.7759/cureus.45075.
56. Wang, K., Wei, Y., Tang, J., Wang, Y., Li, Z., Tong, M., ... & Zhao, Z. (2024, December). Camera-based hrv prediction for remote learning environments. In *2024 IEEE Smart World Congress (SWC)* (pp. 1165-1173). IEEE. https://doi.org/10.1109/SWC62898.2024.00185.
57. Huang, R. Y., & Dung, L. R. (2016). Measurement of heart rate variability using off-the-shelf smart phones. *Biomedical Engineering Online*, *15*(1), 11. https://doi.org/10.1186/s12938-016-0127-8.
58. Odinaev, I., Wong, K. L., Chin, J. W., Goyal, R., Chan, T. T., & So, R. H. (2023). Robust heart rate variability measurement from facial videos. *Bioengineering*, *10*(7), 851. https://doi.org/10.3390/bioengineering10070851.
59. Speth, J.; Vance, N.; Flynn, P.; Bowyer, K.; Czajka, A. Unifying frame rate and temporal dilations for improved remote pulse detection. *Comput. Vis. Image Underst*. 2021, 210, 103246. https://doi.org/10.1016/j.cviu.2021.103246.
60. Bobbia, S.; Macwan, R.; Benezeth, Y.; Mansouri, A.; Dubois, J. Unsupervised skin tissue segmentation for remote photoplethysmography. *Pattern Recognition Letter* 2019, 124, 82–90. https://doi.org/10.1016/j.patrec.2017.10.017.

61. van Putten, L. D., Ahmed, A., & Wegerif, S. (2025). Remote photoplethysmography for contactless pulse rate monitoring: algorithm development and accuracy assessment. *Physiological Measurement*, *46*(11), 115004. https://iopscience.iop.org/article/10.1088/1361-6579/ae1804/meta.
62. Sun, Z., Junttila, J., Tulppo, M., Seppänen, T., & Li, X. (2022). Non-contact atrial fibrillation detection from face videos by learning systolic peaks. *IEEE Journal of Biomedical and Health Informatics*, *26*(9), 4587-4598. https://doi.org/10.1109/JBHI.2022.3193117.
63. Kotzen, K., Charlton, P. H., Landesberg, A., & Behar, J. A. (2021, September). Benchmarking photoplethysmography peak detection algorithms using the electrocardiogram signal as a reference. In *2021 Computing in Cardiology (CinC)* (Vol. 48, pp. 1-4). IEEE. https://doi.org/10.23919/CinC53138.2021.9662889.
64. Boccignone, G., Conte, D., Cuculo, V., d'Amelio, A., Grossi, G., & Lanzarotti, R. (2020). An open framework for remote-PPG methods and their assessment. *Ieee Access*, *8*, 216083-216103. https://doi.org/10.1109/ACCESS.2020.3040936.
65. Spiegelberg, J., & Rusz, J. (2017). Can we use PCA to detect small signals in noisy data?. *Ultramicroscopy*, *172*, 40-46. https://doi.org/10.1016/j.ultramic.2016.10.008.
66. Caroppo, A., Manni, A., Rescio, G., Siciliano, P., & Leone, A. (2024). Vital signs estimation in elderly using camera-based photoplethysmography. *Multimedia Tools and Applications*, *83*(24), 65363-65386. https://doi.org/10.1007/s11042-023-18053-3.
67. Lee, C.; Lee, C.; Fernando, C.; Chow, C.M. Comparison of Apple watch vs KardiaMobile: A tale of two devices. *CJC Open 2022*, 4, 939–945. https://doi.org/10.1016/j.cjco.2022.07.011.
68. Johnson, M. J., Chahal, T., Stinchcombe, A., Mullen, N., Weaver, B., & Bédard, M. (2011). Physiological responses to simulated and on-road driving. *International journal of Psychophysiology*, *81*(3), 203-208. https://doi.org/10.1016/j.ijpsycho.2011.06.012.
69. Bondarenko, M., Menon, C., & Elgendi, M. (2025). The role of face regions in remote photoplethysmography for contactless heart rate monitoring. *npj Digital Medicine*, 8(1), 479. https://doi.org/10.1038/s41746-025-01814-9.
70. Kiddle, A., Barham, H., Wegerif, S., & Petronzio, C. (2023). Dynamic region of interest selection in remote photoplethysmography: Proof-of-concept study. *JMIR Formative Research*, 7, e44575. https://formative.jmir.org/2023/1/e44575.
71. Kim, D. Y., Lee, K., & Sohn, C. B. (2021). Assessment of roi selection for facial video-based rPPG. *Sensors*, 21(23), 7923. https://doi.org/10.3390/s21237923.
72. Premkumar, S., & Hemanth, D. J. (2022, August). Intelligent remote photoplethysmography-based methods for heart rate estimation from face videos: A survey. *Informatics*, 9(3), 57. MDPI. https://doi.org/10.3390/informatics9030057.
73. Soares, S., Ferreira, S., & Couto, A. (2020). Driving simulator experiments to study drowsiness: A systematic review. *Traffic Injury Prevention*, 21(1), 29-37. https://doi.org/10.1080/15389588.2019.1706088.
74. Lohani, M., Payne, B. R., & Strayer, D. L. (2019). A review of psychophysiological measures to assess cognitive states in real-world driving. *Frontiers in Human Neuroscience*, 13, 57. https://doi.org/10.3389/fnhum.2019.00057.
75. Rasche, S., Trumpp, A., Waldow, T., Gaetjen, F., Plötze, K., Wedekind, D., ... & Zaunseder, S. (2016). Camera-based photoplethysmography in critical care patients. *Clinical Hemorheology and Microcirculation*, 64(1), 77-90. https://doi.org/10.3233/CH-162048.
76. Nussinovitch, U., Elishkevitz, K. P., Kaminer, K., Nussinovitch, M., Segev, S., Volovitz, B., & Nussinovitch, N. (2011). The efficiency of 10-second resting heart rate for the evaluation of short-term heart rate variability indices. *Pacing and Clinical Electrophysiology*, *34*(11), 1498-1502. https://doi.org/10.1111/j.1540-8159.2011.03178.x.
77. Developed with the Special Contribution of the European Heart Rhythm Association (EHRA), Endorsed by the European Association for Cardio-Thoracic Surgery (EACTS), Authors/Task Force Members, Camm, A. J., Kirchhof, P., Lip, G. Y., ... & Zupan, I. (2010). Guidelines for the management of atrial fibrillation: the Task Force for the Management of Atrial Fibrillation of the European Society of Cardiology (ESC). *European heart journal*, *31*(19), 2369-2429. https://doi.org/10.1093/eurheartj/ehq278.
78. Debnath, U., & Kim, S. (2026). Advanced signal-processing framework for remote photoplethysmography-based heart rate measurement: Integrating adaptive Kalman filtering with discrete wavelet transformation. *PLoS One*, *21*(1), e0340097. https://doi.org/10.1371/journal.pone.0340097.